\documentclass[twoside,11pt]{article}

\usepackage{blindtext}

\usepackage[preprint]{jmlr2e}

\usepackage{latexsym}
\usepackage{psfrag,epsf}
\usepackage{amsmath}
\usepackage{graphicx}
\usepackage{enumerate}
\usepackage{url} 
\usepackage[utf8]{inputenc}
\usepackage{booktabs}
\usepackage{amssymb}
\usepackage[colorinlistoftodos]{todonotes}
\usepackage{caption}
\usepackage{subcaption}
\usepackage{lscape}
\usepackage{verbatim}
\usepackage{csvsimple}
\usepackage{multirow}
\usepackage{textcomp}
\usepackage{natbib}
\usepackage{mathabx}
\usepackage{tabularx}
\usepackage{bm}
\usepackage{wrapfig}
\usepackage{pdfpages,setspace}
\usepackage[normalem]{ulem}
\usepackage{booktabs}
\usepackage{enumitem}
\usepackage{algorithm}
\usepackage{algpseudocode}
\usepackage{adjustbox}

\usepackage{enumitem}
\usepackage{mathrsfs}

\newtheorem{condition}{Condition}

\newcommand{\E}{\mathbb E}
\newcommand{\Var}{\mathrm Var}
\newcommand{\Pbb}{\mathbb P}

\newcommand{\KL}{\operatorname{KL}}

\newcommand{\logit}{\operatorname{logit}}

\newcommand{\tr}{\operatorname{tr}}
\newcommand{\Cov}{\operatorname{Cov}}

\usepackage{lastpage}
\jmlrheading{23}{2026}{1-\pageref{LastPage}}{1/21; Revised 5/22}{9/22}{21-0000}{Xingche Guo}

\ShortHeadings{Anchored Variational Inference for Personalized Sequential Latent-State Models}{Guo, X.}
\firstpageno{1}

\begin{document}

\title{Anchored Variational Inference for Personalized Sequential Latent-State Models}

\author{\name Xingche Guo \email xingche.guo@uconn.edu \\
       \addr Department of Statistics\\
       University of Connecticut\\
       Storrs, CT 06269, USA
      }

\editor{}

\maketitle

\begin{abstract}
Sequential latent-variable models with  subject-specific random effects provide a flexible framework for modeling temporally structured data with both local latent dynamics and stable between-subject heterogeneity. In such models, conditional inference for the local latent process is often tractable, but integrating over subject-specific random effects can be computationally demanding. We propose an anchored variational inference framework for efficient approximate inference in this setting. The central idea is to replace the full conditional posterior of the local latent process with its evaluation at a representative value of the subject-specific latent effect, called the anchor point, thereby preserving tractable local inference while substantially reducing computational cost. This approximation is especially appealing in sequential settings, where the posterior distribution of the random effect becomes increasingly concentrated as the sequence length grows. 
Under suitable conditions, we show that the posterior mean is a nearly optimal anchor point and that the resulting anchored variational EM (AVEM) algorithm approximately preserves the local monotonicity behavior of standard variational inference. We instantiate the framework in two representative classes of sequential latent-variable models, namely mixed hidden Markov models and mixed-effects state-space models, derive the corresponding AVEM algorithms, and use simulation studies to indicate that the resulting methods achieve accurate estimation with substantial computational gains. We also discuss a partially anchored variant of the framework, in which only the components of the subject-specific latent effect whose posteriors are well concentrated are anchored.
\end{abstract}


\section{Introduction}

Sequential latent-variable models are a central tool for modeling temporally dependent data with latent structure. Classical hidden Markov models (HMMs) provide a flexible framework for regime-switching sequences, while state-space models (SSMs) offer a complementary framework for continuous latent dynamics. These models have been widely used in speech recognition, econometrics, neuroscience, behavioral science, and engineering \citep{shumway1982approach, rabiner1989tutorial, cappe2005inference, ashwood2022mice, guo2025hmm}. 
In many modern applications, however, the inferential challenge lies not only in modeling sequential dependence, but also in accounting for persistent between-subject heterogeneity. Repeated measurements are often collected from multiple subjects or units that exhibit both time-varying latent behavior and stable subject-specific variability. These considerations motivate sequential latent-variable models that combine local latent dynamics with subject-specific random effects.

Sequential latent-variable models with subject-specific random effects arise in a range of applications where one seeks to capture both within-unit temporal dependence and persistent cross-unit variability. Examples include behavioral and cognitive studies, in which subjects may evolve over time through latent decision regimes while differing in stable response tendencies or learning parameters \citep{bian2025ddm}, as well as applications in neuroscience, animal movement, and marketing panel analysis \citep{guo2024hierarchical, mcclintock2021worth, kappe2018random}. These settings have motivated mixed or hierarchical extensions of HMMs and related sequential models that combine subject-specific random effects with latent temporal structure; see, for example, \citet{altman2007mixed}, \citet{maruotti2011mixed}, \citet{coviello2014clustering}, and \citet{mcclintock2021worth}. Although such extensions substantially broaden modeling flexibility, they also introduce significant inferential and computational challenges.

The main computational difficulty in these models is the posterior coupling between the local latent process and the subject-specific continuous latent effect. Conditional on a given value of the subject-specific latent effect, local latent inference may remain tractable: in HMMs one can use forward--backward recursions \citep{baum1970maximization}, while in linear Gaussian SSMs one can use Kalman filtering and smoothing \citep{welch1995introduction}. Once the subject-specific latent effect must itself be integrated out, however, these tractable local inference routines become embedded in a continuous approximation problem. As a result, exact likelihood-based inference is generally unavailable outside relatively simple special cases, and existing approaches often require repeated numerical integration over the random effects.

Variational inference provides a scalable alternative to exact likelihood-based computation by replacing the intractable posterior distribution with a tractable approximating family \citep{neal1998view, jordan1999introduction, ormerod2010explaining, blei2017variational, zhang2018advances}. 
Although fully factorized mean-field approximations \citep{blei2017variational} offer substantial computational convenience, they can be overly restrictive for sequential data because they break the dependence structure among local latent variables.
Such dependence is often intrinsic to the posterior structure; ignoring it may
degrade approximation quality and compromise latent-state inference
\citep{wang2004lack}. More structured variational approximations have therefore been recognized as important for capturing posterior dependence in a broad range of latent-variable models, including sequential and graphical models \citep{saul1995exploiting, ghahramani2000variational, xing2003generalized, foti2014stochastic, ranganath2016hierarchical, wu2026extending}.
Nevertheless, it remains challenging to construct approximations that preserve the essential dependence structure between the local latent process and the subject-specific random effects while retaining computational simplicity.

In this paper, we propose an \emph{anchored variational inference} framework for sequential latent-variable models with tractable conditional local inference and continuous subject-specific latent effects. Our starting point is a structured variational factorization that preserves the conditional dependence of the local latent process on the subject-specific latent effect. Direct optimization over this family remains computationally burdensome, because the conditional local posterior must be recomputed across many values of the subject-specific latent variable. To address this difficulty, we introduce an anchored approximation that replaces the full conditional local posterior with its evaluation at a representative value of the subject-specific latent effect, referred to as the anchor point. This construction preserves the tractable local computational structure of the original model, such as forward--backward inference in HMMs \citep{baum1970maximization}, while avoiding repeated recomputation over the full conditional family.

The effectiveness of the proposed method relies on a basic feature of many sequential settings: as the sequence length increases, the posterior distribution of the subject-specific latent effect often becomes increasingly concentrated around its center. This concentration suggests that a representative anchor point, such as the posterior mean, can provide an accurate approximation to the conditional local posterior. Under suitable conditions, we show that the posterior mean is a nearly optimal anchor point and that the resulting anchored variational EM updates approximately preserve the local monotonicity behavior of standard variational inference.

We instantiate the proposed framework in two representative classes of sequential latent-variable models, namely mixed hidden Markov models \citep[MHMMs;][]{altman2007mixed} and mixed-effects state-space models \citep[MESSMs;][]{liu2011mixed}, and use them as working examples for developing the corresponding estimation procedures. More broadly, the contribution of this paper is an approximate inference framework for models that combine tractable conditional local dynamics with subject-specific latent heterogeneity. Although our focus is on sequential models, the same anchoring principle may also extend to other graphical models with analogous conditional structure \citep{jordan1999introduction}.
We also discuss a partially anchored variant of the framework, in which anchoring is applied only to components of the subject-specific latent effect whose posteriors are well concentrated, while numerical integration is retained for components whose posteriors remain more diffuse. This extension further illustrates the flexibility of the proposed framework in settings with heterogeneous inferential difficulty across latent-effect components.

The remainder of the paper is organized as follows. In Section~\ref{sec:method}, we introduce the personalized sequential latent-state model and the proposed anchored variational inference framework. Section~\ref{sec:theory} provides theoretical justification for the proposed approach. Section~\ref{sec:algorithm} develops the AVEM algorithm for the MHMM case, discusses its relationship to numerically implemented exact EM and structured mean-field variational inference, and presents a Gaussian-emission special case with closed-form updates. Section~\ref{sec:algorithm_ssm} presents the AVEM algorithm for MESSMs. Section~\ref{sec:simulation} reports simulation results, and Section~\ref{subsec:partial_anchor} discusses a partially anchored extension of the proposed framework. Finally, Section~\ref{sec:discussion} concludes with a summary of the main contributions and key findings. Additional methodological details and proofs of the main theoretical results are deferred to the Appendix.

\section{Methodology}
\label{sec:method}

\subsection{Personalized sequential latent-state models}

We consider longitudinal data collected from \(n\) subjects. For subject \(i\), let $D_i=(D_{i1},\dots,D_{iT_i})$
denote the observed sequence of trial-level outcomes, where \(D_{it}\) may be multivariate and may contain components of different types measured at time \(t\).

To characterize within-subject regime-switching dynamics, we introduce a latent state process $U_i=(U_{i1},\dots,U_{iT_i})$,
which is modeled as a first-order Markov process:
\begin{align*}
p(U_i; \omega)
=
p(U_{i1}; \omega)\prod_{t=2}^{T_i}p(U_{it}\mid U_{i,t-1}; \omega).
\end{align*}
Although we focus on Markovian latent processes for concreteness, the proposed framework extends naturally to more general local latent structures, such as hidden semi-Markov models \citep{yu2010hidden} and other graphical models, whenever conditional local inference is tractable.

To capture stable between-subject heterogeneity, we further introduce a subject-specific latent random effect
\[
f_i \sim p(f_i;\eta),
\]
independently across subjects.

Conditional on \(U_i\) and \(f_i\), we assume that the observed sequence \(D_i\) admits the factorization
\[
p(D_i\mid U_i,f_i; \theta_e)
=
\prod_{t=1}^{T_i}
p(D_{it}\mid U_{it},f_i,D_{i[1:t-1]}; \theta_e),
\]
where
$D_{i[1:t-1]}=(D_{i1},\dots,D_{i,t-1})$
denotes the observed history prior to time \(t\). Thus, the conditional distribution of \(D_{it}\) is allowed to depend on the current latent state \(U_{it}\), the subject-specific random effect \(f_i\), and the past observed history.
This formulation is intentionally broad. It includes ordinary HMMs and SSMs as special cases, and also accommodates
autoregressive hidden Markov models \citep{hamilton1989new} and
reinforcement-learning hidden Markov models \citep{guo2025hmm}, in which the emission
distribution depends on history-dependent features or recursively updated covariates.
In more general settings, one may also allow the latent state process \(U_i\) to depend on \(f_i\). For simplicity, we do not impose that additional dependence here.

\subsection{Anchored variational inference}

To estimate the model parameters $\theta = (\theta_e, \eta, \omega)$, we need to maximize the observed-data likelihood, which requires integrating out the subject-specific latent variable \(f_i\) and marginalizing over the latent process \(U_i\). In most sequential latent-state models, this calculation is analytically intractable. For example, in an HMM with \(T\) time points and \(K\) latent states, a brute-force evaluation would require summing over \(K^T\) possible latent state sequences.

A classical alternative is the EM algorithm \citep{dempster1977maximum}, which is based on the complete-data log-likelihood
\begin{align}
\log p(D_i,U_i,f_i;\theta)
&=
\log p(f_i;\eta)
+
\log p(U_{i1};\omega)
+
\sum_{t=2}^{T_i}\log p(U_{it}\mid U_{i,t-1};\omega)
\notag\\
&\quad+
\sum_{t=1}^{T_i}\log p(D_{it}\mid U_{it},f_i,D_{i[1:t-1]};\theta_e).
\label{eq:complete_loglik_full}
\end{align}

However, its E-step requires computing the posterior distribution $p(U_i,f_i\mid D_i)$,
which is generally also intractable in the models considered here.

A natural approach is therefore to use variational inference. Specifically, for each subject \(i\), we introduce a variational distribution \(q_i(U_i,f_i)\) to approximate the intractable posterior \(p(U_i,f_i\mid D_i)\). Let $q=(q_1,\dots,q_n)$
denote the collection of subject-specific variational distributions. We then estimate \(\theta\) by maximizing the evidence lower bound (ELBO),
\[
\mathcal L(q,\theta)
=
\sum_{i=1}^n
\E_{q_i(U_i,f_i)}
\left[
\log p(D_i,U_i,f_i;\theta)
-
\log q_i(U_i,f_i)
\right].
\]

It is well known that the evidence lower bound is maximized, or equivalently that the Kullback--Leibler divergence from \(q_i(U_i,f_i)\) to the true posterior is minimized, when
\[
q_i(U_i,f_i)=p(U_i,f_i\mid D_i).
\]
In this case, the variational E-step reduces to the exact E-step in the EM algorithm \citep{neal1998view}.

For subject \(i\), we may decompose the complete posterior as
\[
p(U_i,f_i\mid D_i)
=
p(U_i\mid f_i,D_i)\,p(f_i\mid D_i),
\]
where \(p(U_i\mid f_i,D_i)\) denotes the conditional posterior distribution of the latent process \(U_i\) given \(f_i\) and the observed data \(D_i\).
In many important sequential latent-state models, conditioning on \(f_i\) restores a standard latent Markov structure, so \(p(U_i\mid f_i,D_i)\) can be computed efficiently using model-specific smoothing algorithms, such as the forward--backward algorithm for HMMs \citep{baum1970maximization} or the Kalman smoother for SSMs \citep{welch1995introduction}.
However, the marginal posterior \(p(f_i\mid D_i)\) is generally intractable or computationally expensive to compute; see Section \ref{sec:exact_em_vs_avem} and the simulation studies for more details.

A natural variational approximation is to adopt the \emph{structured factorization}
\begin{align}
    q_i(U_i,f_i)=q_i(U_i\mid f_i)\,q_i(f_i), \label{equ:structured_factorization}
\end{align}
with \(q_i(f_i)\) taken from a simple variational family, such as a Gaussian distribution. This formulation is appealing because it imposes an approximation only on the marginal posterior of \(f_i\), while preserving flexibility in the conditional distribution of the latent process \(U_i\) given \(f_i\). Indeed, once \(q_i(f_i)\) is fixed, Lemma~\ref{lemma1} implies that the optimal conditional factor is
\[
q_i(U_i\mid f_i)=p(U_i\mid f_i,D_i).
\]
Hence, the resulting variational family preserves the exact conditional posterior \(p(U_i\mid f_i,D_i)\) and approximates only the marginal posterior \(p(f_i\mid D_i)\).

\begin{lemma} \label{lemma1}
Let $q_i(U_i,f_i)=q_i(U_i \mid f_i)\,q_i(f_i)$ and $p(U_i,f_i \mid D_i)=p(U_i\mid f_i, D_i)\,p(f_i \mid D_i)$.
If \(q_i(f_i)\) is fixed, then \(\KL\bigl(q_i(U_i,f_i)\|p(U_i,f_i\mid D_i)\bigr)\) is minimized when $q_i(U_i\mid f_i)=p(U_i\mid f_i,D_i)$.
\end{lemma}

This formulation is attractive because it implies that only the marginal variational factor \(q_i(f_i)\) needs to be approximated. Nevertheless, directly implementing this structured factorization remains computationally demanding in an iterative algorithm. Indeed, each update requires integration with respect to \(q_i(f_i)\), and for each value of \(f_i\), the conditional posterior $p(U_i\mid f_i,D_i)$
must be recomputed. Consequently, despite preserving the exact conditional form of the latent-state posterior, the resulting variational updates can still be prohibitively expensive.

This observation motivates a compromise between the fully structured factorization in \eqref{equ:structured_factorization} and a more computationally tractable approximation. To this end, we introduce an \emph{anchored family}, which preserves part of the dependence between \(U_i\) and \(f_i\) while avoiding the full computational burden of the structured formulation.

\begin{definition}[Anchor family]
For subject \(i\), define the anchor variational family by
\[
q_i(U_i,f_i)
=
p(U_i\mid f_{0i},D_i)\,q_i(f_i),
\]
where \(f_{0i}\) is a fixed anchor point.
\end{definition}

The anchor \(f_{0i}\) serves as a representative value of \(f_i\), at which the conditional posterior \(p(U_i\mid f_i,D_i)\) is evaluated and then held fixed, rather than allowed to vary with \(f_i\).

A natural question is how to choose a good anchor \(f_{0i}\). 
Lemma~\ref{lemma2} below characterizes the optimal choice of anchor.

\begin{lemma} \label{lemma2}
Suppose \(q_i(U_i,f_i)=q_i(U_i\mid f_i)\,q_i(f_i)\). If $q_i(f_i)$ is fixed and we impose the restriction \(q_i(U_i\mid f_i)=p(U_i\mid f_{0i},D_i)\) for some fixed \(f_{0i}\). Then \(\KL\bigl(q(U_i,f_i)\|p(U_i,f_i\mid D_i)\bigr)\) is minimized when 
\[
f_{0i}^\ast
=
\arg\min_{f_{0}}\;
\mathbb E_{q(f_i)}
\!\left[
\KL\bigl(p(U_i\mid f_{0},D_i)\|p(U_i\mid f_i,D_i)\bigr)
\right].
\]
\end{lemma}

Lemma~\ref{lemma2} characterizes the optimal anchor point. However, under a variational EM algorithm, the model parameters \(\theta\) are updated at each iteration, and hence the optimal anchor point may vary across iterations as well. As a result, recomputing the optimal anchor point at every step can be computationally costly, since it requires solving an additional numerical optimization problem within each iteration.

A simple and appealing choice is the \emph{mean anchor}, which replaces the optimal anchor by the mean of the current variational distribution of \(f_i\). This avoids the extra numerical optimization required by the optimal-anchor strategy.

\begin{definition}[Mean anchor]
For subject \(i\), define the mean-anchor variational distribution by
\[
q_i(U_i,f_i)
=
p(U_i\mid \bar f_{0i},D_i)\,q_i(f_i),
\]
where
\[
\bar f_{0i}=\E_{q_i(f_i)}(f_i).
\]
In implementation, \(\bar f_{0i}\) is taken from the previous iteration.
\end{definition}

When \(q_i(f_i)\) is Gaussian, \(\bar f_{0i}\) is immediately available from its variational parameters, so the mean anchor incurs essentially no additional computational cost. It can therefore be viewed as a simple and practically efficient surrogate for the optimal anchor.

\begin{remark}
Anchored variational inference bears some resemblance to Gibbs sampling in a fully Bayesian setting. Both approaches exploit conditional distributions obtained by holding one block fixed. In Gibbs sampling, one alternates between
\[
U_i^{(m+1)} \sim p(U_i \mid f_i^{(m)}, D_i)
\quad \text{and} \quad
f_i^{(m+1)} \sim p(f_i \mid U_i^{(m+1)}, D_i).
\]
Anchored variational inference, however, serves a different purpose. Rather than generating samples from the posterior distribution, it constructs a tractable approximation by anchoring the conditional distribution of \(U_i\) at a fixed value \(f_{0i}\), that is, by using \(p(U_i \mid f_{0i}, D_i)\) in place of \(p(U_i \mid f_i, D_i)\).
\end{remark}

\begin{remark}
It is also natural to consider a blocked mean-field variational family of the form
\begin{align}
q_i(U_i,f_i)=q_i(U_i)\,q_i(f_i), 
\label{equ:block_MF}
\end{align}
which preserves the Markov dependence within \(U_i\) but breaks the posterior dependence between \(U_i\) and \(f_i\). 
As discussed in Section~\ref{sec:block_MFVI}, the proposed anchored update is closely related to the corresponding blocked variational update for \(q_i(U_i)\), while remaining distinct in its use of an anchor point to preserve tractable conditional local inference. This yields a particularly simple and computationally efficient implementation relative to the blocked mean-field alternative.
\end{remark}

\section{Theoretical Properties}
\label{sec:theory}

For simplicity, throughout this section we assume that \(T_i\equiv T\) for all \(i=1,\dots,n\).
We begin by quantifying the approximation error of the anchor family relative to the fully structured variational formulation.

\begin{condition}[KL-smoothness of the conditional posterior]
\label{cond:KL_smooth}
Assume that, for $\theta$ in a neighborhood of the true parameter value $\theta^\ast$, 
the conditional posterior family
\[
\{\,p(U_i\mid D_i,f;\theta): f\in\mathbb R^d\,\}
\]
is KL-smooth in $f$. Specifically, assume that there exists a constant $L>0$ such that, 
almost surely with respect to the distribution of $D_i$,
\begin{align}
\KL\!\left(
p(U_i\mid D_i,f_0;\theta)
\,\big\|\,
p(U_i\mid D_i,f;\theta)
\right)
\le
L T \|f_0-f\|^2
\label{eq:cond_KL_smooth}
\end{align}
for all $f_0,f\in\mathbb R^d$ and all $i=1,\dots,n$.
\end{condition}

\begin{remark}
A useful way to interpret Condition~\ref{cond:KL_smooth} is through a local quadratic approximation.
Define
\[
G_i(f;f_0,D_i,\theta)
:=
\KL\!\left(
p(U_i\mid D_i,f_0;\theta)
\,\big\|\,
p(U_i\mid D_i,f;\theta)
\right).
\]
Because $G_i(f_0;f_0,D_i,\theta)=0$,
and \(f=f_0\) is a minimizer of \(G_i(\,\cdot\,;f_0,D_i,\theta)\). Hence, assuming differentiability with respect to \(f\), the first-order condition implies that
\[
\nabla_f G_i(f_0;f_0,D_i,\theta)=0.
\]
Under sufficient differentiability, a second-order Taylor expansion of \(G_i(f;f_0,D_i,\theta)\) at \(f=f_0\) yields
\[
G_i(f;f_0,D_i,\theta)
=
\frac12 (f-f_0)^\top \nabla_f^2 G_i(\tilde f;f_0,D_i,\theta)(f-f_0),
\]
for some \(\tilde f\) lying on the line segment between \(f_0\) and \(f\). Consequently, if
\[
\lambda_{\max}\!\left\{\nabla_f^2 G_i(\tilde f;f_0,D_i,\theta)\right\}
\le 2LT
\]
uniformly over the relevant range of \(f\), then \eqref{eq:cond_KL_smooth} holds. The linear scaling in \(T\) reflects the fact that, in sequential latent-variable models, posterior information and curvature often accumulate roughly additively across time points.

\end{remark}

\begin{theorem}[\textbf{Near-optimality of the mean anchor}]
\label{thm:mean_anchor_bound}
Suppose Condition~\ref{cond:KL_smooth} holds, and define
\[
\mathcal J_i(f_{0i})
:=
\mathbb E_{q_i(f_i)}
\!\left[
\KL\!\left(
p(U_i\mid D_i,f_{0i};\theta)
\,\big\|\,
p(U_i\mid D_i,f_i;\theta)
\right)
\right].
\]
Let $\bar f_{0i}:=\mathbb E_{q_i(f_i)}(f_i)$
denote the mean anchor, and let
$f_{0i}^\ast:=\arg\min_{f_{0i}}\mathcal J_i(f_{0i})$
denote the optimal anchor. Then
\[
\frac{1}{T} \left\{\mathcal J_i(\bar f_{0i})-\mathcal J_i(f_{0i}^\ast) \right\}
\le
L\tr\bigl(\Var_{q_i}(f_i)\bigr)
\qquad a.s.
\]
\end{theorem}

Theorem~\ref{thm:mean_anchor_bound} bounds the normalized suboptimality of the mean anchor in terms of the concentration of \(q_i(f_i)\). This concentration is generally not expected in ordinary non-sequential settings. In the present sequential setting, however, each subject typically provides a relatively long trajectory, so it is reasonable to expect that the posterior uncertainty about the subject-specific random effect \(f_i\) shrinks as \(T\) increases. The following condition formalizes this requirement.

\begin{condition}\label{cond:var_concentration}
For some \(0<r\le 1\), assume that there exists a constant \(L'>0\) such that
\[
\max_{i\in \{1, \dots, n\}}\tr\bigl(\Var_{q_i}(f_i)\bigr) \le L' T^{-r}
\qquad \text{a.s.}
\]
\end{condition}

\begin{remark}
Condition~\ref{cond:var_concentration} is motivated by the fact that, for each subject, the amount of information about \(f_i\) typically increases with the sequence length \(T\). In particular models, such as the mixture hidden Markov model with Gaussian emissions considered in Section~\ref{sec:special_case}, the precision is shown to increase linearly with \(T\), which in turn yields a variance of order \(1/T\).
Condition~\ref{cond:var_concentration} imposes a weaker requirement by only assuming a rate \(T^{-r}\) for some \(0<r\le 1\), allowing for slower concentration in more complex or weakly identified models.
\end{remark}

\begin{corollary}\label{cor:mean_anchor_rate}
Suppose Conditions~\ref{cond:KL_smooth} and \ref{cond:var_concentration} hold. Then
\[
\frac{1}{T}\left\{\mathcal J_i(\bar f_{0i})-\mathcal J_i(f_{0i}^\ast)\right\}
\le
LL' T^{-r} \qquad a.s.
\]
In particular, the normalized suboptimality of the mean anchor vanishes as \(T\to\infty\).
\end{corollary}

Next, we study the approximate monotonicity property of the anchored variational EM algorithm.

For the anchor variational family, let \(q=(q_1,\dots,q_n)\) denote the collection of variational distributions and let \(f_0=(f_{01},\dots,f_{0n})\) denote the collection of anchor points. The corresponding anchored ELBO is
\begin{align}
\mathcal L_A(f_0,q,\theta)
&=
\sum_{i=1}^n
\mathbb E_{q_i(f_i)\,p(U_i\mid D_i,f_{0i};\theta)}
\Bigl[
\log p(D_i,U_i,f_i;\theta)
-
\log p(U_i\mid D_i,f_{0i};\theta) \notag \\
& \qquad \qquad \qquad \qquad \qquad \qquad
-
\log q_i(f_i)
\Bigr].
\label{equ:anchor_elbo}
\end{align}

Since \(\mathcal L_A(f_0,q,\theta)\) aggregates contributions over all \(n\) subjects and \(T\) time points, its magnitude is naturally of order \(nT\). Hence, it is more meaningful to work with the normalized anchored ELBO,
\[
\bar{\mathcal L}_A(f_0,q,\theta)
=
\frac{1}{nT}\mathcal L_A(f_0,q,\theta).
\]

The following proposition shows that, when the anchor is fixed, the proposed updates are monotone with respect to the anchored ELBO.

\begin{proposition}
\label{prop:fixed_anchor_monotonicity}
Fix \(f_0\). Suppose that, at iteration \(m\), the updates for \(p(U\mid D, f_0)\), \(q(f)\), and \(\theta\) are obtained by exact blockwise maximization of the anchored ELBO in \eqref{equ:anchor_elbo}. Then
\[
\bar{\mathcal L}_A\!\left(f_0,q^{(m+1)},\theta^{(m+1)}\right)
\ge
\bar{\mathcal L}_A\!\left(f_0,q^{(m)},\theta^{(m)}\right).
\]
\end{proposition}

Proposition \ref{prop:fixed_anchor_monotonicity} is immediate because, for fixed \(f_0\), the anchored ELBO reduces to an ordinary ELBO over a restricted variational family. Therefore, exact blockwise maximization yields a nondecreasing coordinate-ascent procedure \citep{bertsekas1997nonlinear}.

The next theorem quantifies the effect of refreshing the anchor point across iterations and identifies conditions under which approximate monotonicity holds.

\begin{theorem}[\textbf{Asymptotic Monotonicity}]
\label{thm:approx_monotonicity}
Suppose Conditions~\ref{cond:KL_smooth}--\ref{cond:var_concentration} hold. In addition, suppose that \(\theta\) lies in a neighborhood of the true parameter value \(\theta^\ast\), and that, at iteration \(m\), the anchor is chosen as the mean-anchor.
Then, for all sufficiently large \(T\),
\[
\bar{\mathcal L}_A\!\left(f_0^{(m+1)},q^{(m+1)},\theta^{(m+1)}\right)
\ge
\bar{\mathcal L}_A\!\left(f_0^{(m)},q^{(m)},\theta^{(m)}\right)
-
L L' T^{-r} \qquad a.s.
\]
\end{theorem}

Theorem~\ref{thm:approx_monotonicity} establishes that the normalized anchored ELBO is approximately nondecreasing across iterations, up to an error of order \(O(T^{-r})\).

Let \(\mathcal L_S(q,\theta)\) denote the ELBO associated with the structured factorization in \eqref{equ:structured_factorization}. The following result compares the anchored ELBO \(\mathcal L_A(f_0,q,\theta)\) with its structured counterpart \(\mathcal L_S(q,\theta)\).

\begin{theorem}
\label{thm:elbo_gap}
Suppose Conditions~\ref{cond:KL_smooth}--\ref{cond:var_concentration} hold. Let \(\mathcal L_S(q,\theta)\) be the ELBO corresponding to the structured factorization in \eqref{equ:structured_factorization}, Further define $\bar{\mathcal L}_S(q,\theta):=(1/nT)\mathcal L_S(q,\theta)$.
If the anchor is chosen as the mean anchor, then
\[
0 \le \bar{\mathcal L}_S(q,\theta)-\bar{\mathcal L}_A(f_0,q,\theta) \le L L' T^{-r} \qquad a.s.
\]
\end{theorem}

\begin{corollary}
\label{cor:true_elbo_approx_monotonicity}
Suppose Conditions~\ref{cond:KL_smooth}--\ref{cond:var_concentration} hold, then
\[
\bar{\mathcal L}_S\!\left(q^{(m+1)},\theta^{(m+1)}\right)
\ge
\bar{\mathcal L}_S\!\left(q^{(m)},\theta^{(m)}\right)
-
2 L L' T^{-r} \qquad a.s.
\]
In other words, the structured ELBO is approximately nondecreasing along the anchored variational EM iterations, up to an error of order \(O(T^{-r})\).
\end{corollary}

Corollary~\ref{cor:true_elbo_approx_monotonicity} is an immediate consequence of Theorems~\ref{thm:approx_monotonicity} and \ref{thm:elbo_gap}. It shows that the proposed anchored variational EM algorithm approximately inherits the ascent property of standard variational EM: although the algorithm optimizes the anchored ELBO rather than the structured ELBO, the structured ELBO decreases by at most \(O(T^{-r})\) in each iteration. For large \(T\), this per-iteration discrepancy becomes negligible, so the algorithm behaves similarly to a standard ascent-based variational EM procedure at the local stepwise level.

Consequently, for sufficiently large \(T\), the anchored variational EM algorithm may be viewed as an approximate ascent procedure for the normalized structured ELBO. A rigorous guarantee that the final AVEM solution is close to a local maximizer of the structured variational objective would require additional control on the accumulated approximation error. In practice, however, if the algorithm converges in relatively few iterations, then this cumulative error is expected to remain small, suggesting that the solution produced by AVEM should lie close to a local maximizer of the structured variational objective.

\section{Anchored Variational EM for Mixed Hidden Markov Models}
\label{sec:algorithm}

We now illustrate the proposed Anchored Variational EM (AVEM) algorithm in an important special case, namely, the mixed hidden Markov model (MHMM).

In this setting, the latent process \(U_i=(U_{i1},\dots,U_{iT_i})\) is discrete, with
$U_{it}\in\{1,\dots,K\}$,
where \(K\) denotes the number of latent states. We assume that \(U_i\) follows a first-order Markov chain with initial distribution
\[
\Pbb(U_{i1}=k)=\pi_k,
\]
and transition probabilities
\[
\Pbb(U_{it}=\ell \mid U_{i,t-1}=k)=\Gamma_{k\ell},
\qquad k,\ell=1,\dots,K.
\]
Assume that the subject-specific random effect \(f_i \in \mathbb{R}^d \) follows \(N(0,\Sigma)\), where \(\Sigma\) is an unknown positive definite covariance matrix.

Under the anchored variational framework described in the previous section, we further specify the variational factor for \(f_i\) as a Gaussian distribution $q_i(f_i)=N(\nu_i,\Omega_i).$
Accordingly, the anchored variational family takes the form
\[
q_i(U_i,f_i)
=
p(U_i\mid f_{0i},D_i)\,N(f_i\mid \nu_i,\Omega_i).
\]

For \(t=1,\dots,T_i\) and \(k=1,\dots,K\), let
\begin{align}
e_{ikt}(\theta,f_i)
=
p(D_{it}\mid U_{it}=k,f_i,D_{i[1:t-1]};\theta)
\label{equ:emission}
\end{align}
denote the emission density at time \(t\) under latent state \(k\).
In addition, define
\[
\zeta_{ikt}(\theta,f_i)
=
p(U_{it}=k\mid D_i,f_i;\theta),
\qquad
\xi_{ik\ell t}(\theta,f_i)
=
p(U_{it}=k,U_{i,t+1}=\ell\mid D_i,f_i;\theta),
\]
which represent the posterior state probability and the posterior pairwise transition probability, respectively. These quantities can be computed efficiently using the forward--backward algorithm; see the Appendix \ref{sec:forward_backward} for details.

The AVEM algorithm is summarized in Algorithm~\ref{alg:avem_mhmm}; the details of its individual steps are provided in the following subsections.

\subsection{Variational E-step}

At iteration \(m\), the anchored variational approximation takes the form
\[
q_i(U_i,f_i;\theta^{(m-1)})
=
p\!\left(U_i\mid f_{0i}^{(m-1)},D_i;\theta^{(m-1)}\right)\,q_i(f_i),
\]
where the anchor is chosen as the mean anchor
\[
f_{0i}^{(m-1)}=\nu_i^{(m-1)}.
\]
Accordingly, evaluating the conditional factor reduces to computing
\[
\zeta_{ikt}^{(m)}
=
\zeta_{ikt}\!\left(\theta^{(m-1)},f_{0i}^{(m-1)}\right),
\qquad
\xi_{ik\ell t}^{(m)}
=
\xi_{ik\ell t}\!\left(\theta^{(m-1)},f_{0i}^{(m-1)}\right),
\]
which can be obtained from a forward--backward recursion.
The corresponding ELBO can be written as
\begin{align}
\mathcal L^{(m)}(q,\theta)
&=
\sum_{i=1}^n
\sum_{t=1}^{T_i}\sum_{k=1}^K
\zeta_{ikt}^{(m)}
\E_{q_i(f_i)} \Big[ \log e_{ikt}(\theta,f_i) \Big] 
\notag\\
&\qquad
+
\sum_{i=1}^n \Biggl[ \sum_{k=1}^K \zeta_{ik1}^{(m)}\log \pi_k
+
\sum_{t=1}^{T_i-1}\sum_{k=1}^K\sum_{\ell=1}^K
\xi_{ik\ell t}^{(m)}\log \Gamma_{k\ell}
\Biggr]
\notag\\
&\qquad
+
\sum_{i=1}^n \E_{q_i(f_i)}\!\left[\log p\!\left(f_i;\Sigma\right) - \log q_i(f_i)\right]
+
C^{(m)},
\label{eq:elbo_expand_zeta_m_const}
\end{align}
with
\[
C^{(m)}
=
-
\sum_{i=1}^n
\E_{p(U_i\mid f_{0i}^{(m-1)},D_i;\theta^{(m-1)})}
\!\left[
\log p\!\left(U_i\mid f_{0i}^{(m-1)},D_i;\theta^{(m-1)}\right)
\right],
\]
which is constant with respect to $(q, \theta)$.

We can further simplify
\begin{align*}
\E_{q_i(f_i)}\!\left[\log p(f_i;\Sigma)-\log q_i(f_i)\right] 
&=-\frac12\left\{
\operatorname{tr}\!\left(\Sigma^{-1}\Omega_i\right)
-d
+
\nu_i^\top \Sigma^{-1}\nu_i
+
\log\frac{\det(\Sigma)}{\det(\Omega_i)}
\right\}.
\end{align*}

To update the variational distribution \(q_i(f_i)=N(\nu_i,\Omega_i)\), we retain only the terms in the anchored ELBO that depend on \(q_i(f_i)\). This gives
\begin{align*}
\mathcal L_i^{(m)}(\nu_i,\Omega_i)
&=
\sum_{t=1}^{T_i}\sum_{k=1}^K
\zeta_{ikt}^{(m)}
\E_{N(\nu_i,\Omega_i)}
\left[
\log e_{ikt}(\theta^{(m-1)},f_i)
\right]
\notag\\
&\qquad
-\frac12
\left\{
\operatorname{tr}\!\left((\Sigma^{(m-1)})^{-1}\Omega_i\right)
+
\nu_i^\top(\Sigma^{(m-1)})^{-1}\nu_i
-
\log \det(\Omega_i)
\right\}.
\end{align*}
Therefore, the Gaussian variational update is obtained by solving
\begin{align}
    (\nu_i^{(m)},\Omega_i^{(m)})
=
\arg\max_{\nu_i,\Omega_i\succ 0}
\mathcal L_i^{(m)}(\nu_i,\Omega_i).
\label{eq:optimize_q_update}
\end{align}
In general, the expectations in
\(\mathcal L_i^{(m)}(\nu_i,\Omega_i)\) do not admit closed-form expressions. We therefore evaluate them numerically, for example using Gaussian quadrature.

An alternative is to use a Laplace approximation. Under the anchored variational family, the coordinate-wise optimal density for \(f_i\) satisfies
\begin{align*}
q_i^{(m)}(f_i)
\propto
\exp\left\{
\ell_i^A(f_i\mid \theta^{(m-1)},f_{0i}^{(m-1)})
\right\},
\end{align*}
where
\begin{align*}
\ell_i^A(f_i\mid \theta,f_{0i})
&=
\log p(f_i;\theta)
+
\E_{p(U_i\mid D_i,f_{0i};\theta)}
\left[
\log p(D_i,U_i\mid f_i;\theta)
\right]
\notag\\
&=
-\frac12 f_i^\top \Sigma^{-1} f_i
+
\sum_{t=1}^{T_i}\sum_{k=1}^K
\zeta_{ikt}(\theta,f_{0i})
\log e_{ikt}(\theta,f_i)
+
\mathrm{const.},
\end{align*}
and the constant does not depend on \(f_i\).

The Laplace approximation replaces \(q_i^{(m)}(f_i)\) by a Gaussian distribution determined by the mode and local curvature of \(\ell_i^A\). Specifically,
\begin{align}
\nu_i^{(m)}
&=
\arg\max_{f_i\in\mathbb R^d}
\ell_i^A(f_i\mid \theta^{(m-1)},f_{0i}^{(m-1)}),
\notag\\
\Omega_i^{(m)}
&=
\left[
-\nabla^2_{f_i}
\ell_i^A(f_i\mid \theta^{(m-1)},f_{0i}^{(m-1)})
\bigg|_{f_i=\nu_i^{(m)}}
\right]^{-1}.
\label{eq:laplace_q_update}
\end{align}

\subsection{Variational M-step}

We partition the parameter vector as $\theta=(\pi,\Gamma,\theta_e,\Sigma),$
where \(\pi=(\pi_1,\dots,\pi_K)^\top\) is the initial-state distribution, \(\Gamma=(\Gamma_{k\ell})\) is the transition matrix, and \(\theta_e\) collects the global parameters entering the emission model. The M-step then separates naturally into closed-form updates for \((\pi,\Gamma)\) and a numerical optimization for \(\theta_e\).

\paragraph{Update for the initial distribution.}
The part of \eqref{eq:elbo_expand_zeta_m_const} involving \(\pi\) is
\[
\mathcal L_\pi^{(m)}(\pi)
=
\sum_{i=1}^n \sum_{k=1}^K \zeta_{ik1}^{(m)} \log \pi_k,
\qquad
\text{subject to } \sum_{k=1}^K \pi_k=1.
\]
The maximizer is
\begin{align}
\pi_k^{(m)}
=
\frac{1}{n}\sum_{i=1}^n \zeta_{ik1}^{(m)},
\qquad k=1,\dots,K.
\label{eq:update_pi}
\end{align}

\paragraph{Update for the transition matrix.}
The part of \eqref{eq:elbo_expand_zeta_m_const} involving \(\Gamma\) is
\[
\mathcal L_\Gamma^{(m)}(\Gamma)
=
\sum_{i=1}^n \sum_{t=1}^{T_i-1}\sum_{k=1}^K\sum_{\ell=1}^K
\xi_{ik\ell t}^{(m)} \log \Gamma_{k\ell},
\qquad
\text{subject to } \sum_{\ell=1}^K \Gamma_{k\ell}=1,\quad k=1,\dots,K.
\]
Again by Lagrange multipliers, the maximizer is
\begin{align}
\Gamma_{k\ell}^{(m)}
 = \frac{
\sum_{i=1}^n \sum_{t=1}^{T_i-1} \xi_{ik\ell t}^{(m)}
}{
\sum_{i=1}^n \sum_{t=1}^{T_i-1} \zeta_{ikt}^{(m)}
},
\qquad k,\ell=1,\dots,K.
\label{eq:update_gamma}
\end{align}

\paragraph{Update for the emission global parameters.}
Let \(\theta_e\) denote the collection of global parameters appearing in the emission density \(e_{ikt}(\theta,f_i)\). 
The objective function for updating \(\theta_e\) is
\begin{align*}
\mathcal L_e^{(m)}(\theta_e)
:=
\sum_{i=1}^n
\sum_{t=1}^{T_i}\sum_{k=1}^K
\zeta_{ikt}^{(m)}
\E_{q_i^{(m)}(f_i)}
\Big[
\log e_{ikt}(\theta_e,f_i)
\Big].
\end{align*}
Thus, the emission parameters are updated by
\begin{align}
\theta_e^{(m)}
=
\arg\max_{\theta_e} \mathcal L_e^{(m)}(\theta_e).
\label{eq:update_theta_e}
\end{align}
In general, this maximization need not admit a closed-form solution. We
approximate the expectation in \(\mathcal L_e^{(m)}(\theta_e)\) by Gaussian
quadrature or Monte Carlo integration and optimize the resulting objective
numerically. For large-scale problems, one may instead use a generalized EM
update based on a few gradient-ascent steps, possibly with mini-batch
stochastic gradients \citep{bottou2010large}, in the spirit of black-box variational inference \citep{ranganath2014black}.

\paragraph{Update for \(\Sigma\).}
The relevant part of \eqref{eq:elbo_expand_zeta_m_const} is
\[
\mathcal L_\Sigma^{(m)}(\Sigma)
=
-\frac12\sum_{i=1}^n \left\{
\operatorname{tr}\!\left(\Sigma^{-1}\Omega_i\right)
+
\nu_i^\top \Sigma^{-1}\nu_i
+
\log \det(\Sigma)
\right\}.
\]
The maximizer is
\begin{align}
\Sigma^{(m)}
=
\frac{1}{n}
\sum_{i=1}^n
\left\{
\Omega_i^{(m)}+\nu_i^{(m)}\bigl(\nu_i^{(m)}\bigr)^\top
\right\}.
\label{eq:update_sigma}
\end{align}

\begin{algorithm}[ht]
\caption{AVEM Algorithm for MHMMs}
\label{alg:avem_mhmm}
\begin{algorithmic}[1]
\Require Data $\{D_i\}_{i=1}^n$, initialize $r\gets 0$, $\theta^{(0)}$, $\{\nu_i^{(0)},\Omega_i^{(0)}\}_{i=1}^n$

\Repeat
\State \textbf{E-step:}
    \For{$i=1,\dots,n$}      
        \State Let $f_{0i}^{(r)}= \nu_i^{(r)}$ and update
        $p(U_i \mid D_i, f_{0i}^{(r)}; \theta^{(r)})$ via forward-backward algorithm;
        
        \State Update $q_i(f_i)=N(\nu_i^{(r+1)},\Omega_i^{(r+1)})$
        by solving  \eqref{eq:optimize_q_update} or \eqref{eq:laplace_q_update};
    \EndFor
    
    \State \textbf{M-step:} update $\theta^{(r+1)}$ by maximizing the anchored ELBO with maximizer \eqref{eq:update_pi}-\eqref{eq:update_sigma}
    \State $r \gets r+1$
\Until{relative ELBO convergence or maximum iterations reached}

\State \Return $\widehat{\theta}=\theta^{(r)}$
\end{algorithmic}
\end{algorithm}

\subsection{AVEM versus numerically implemented exact EM.} \label{sec:exact_em_vs_avem}

For the MHMM special case, the exact EM algorithm \citep{altman2007mixed, maruotti2011mixed} can in principle be implemented by numerically evaluating the posterior distribution of the random effect,
\[
p(f_i \mid D_i;\theta)
=
\frac{p(D_i \mid f_i;\theta)\,p(f_i;\theta)}
{\int p(D_i \mid f;\theta)\,p(f;\theta)\,df}.
\]
For any fixed \(f_i\), the conditional likelihood \(p(D_i \mid f_i;\theta)\) can be computed using the forward--backward algorithm. Consequently, the exact E-step may be carried out by repeatedly evaluating the forward--backward recursion over a collection of quadrature nodes or Monte Carlo samples, and then numerically approximating the required posterior expectations.

When \(T_i\) is large, the posterior distribution \(p(f_i \mid D_i)\) is expected to become increasingly concentrated, since the subject-specific random effect \(f_i\) is informed by a larger number of repeated observations. As \(p(f_i \mid D_i)\) becomes narrower and more sharply peaked, direct numerical approximation of the exact E-step becomes increasingly challenging. In Monte Carlo EM (MCEM), one typically draws samples from the prior \(p(f_i;\theta)\) and reweights them according to the likelihood term \(p(D_i\mid f_i;\theta)\), so a large number of Monte Carlo samples may be needed in order to adequately capture the small region where the posterior mass is concentrated. Similarly, Gaussian--Hermite quadrature relies on a fixed collection of prior-centered nodes and may therefore require a large number of quadrature points to accurately resolve a posterior that is highly concentrated or shifted away from the prior mean.

\begin{figure}[ht]
    \centering
    \includegraphics[width=0.47\textwidth]{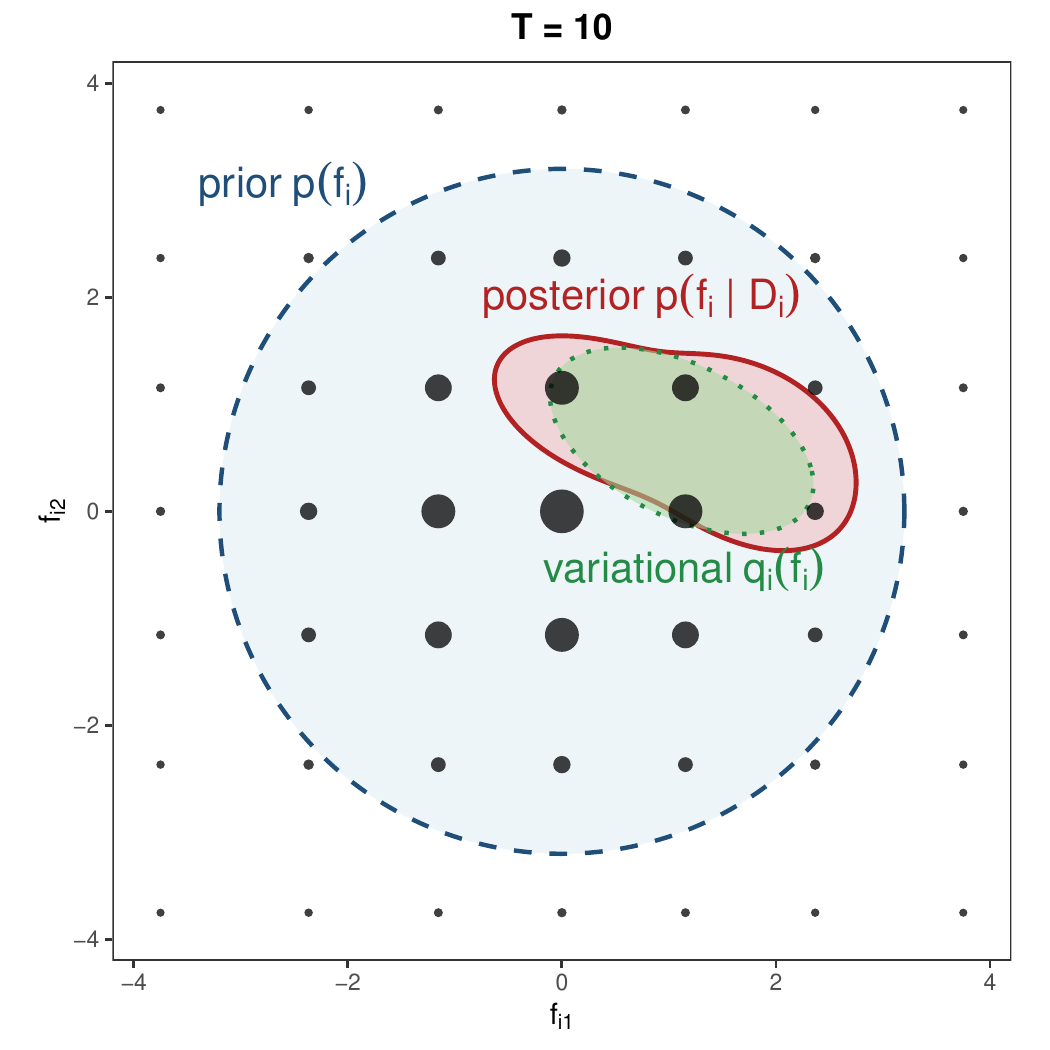}\hfill
    \includegraphics[width=0.47\textwidth]{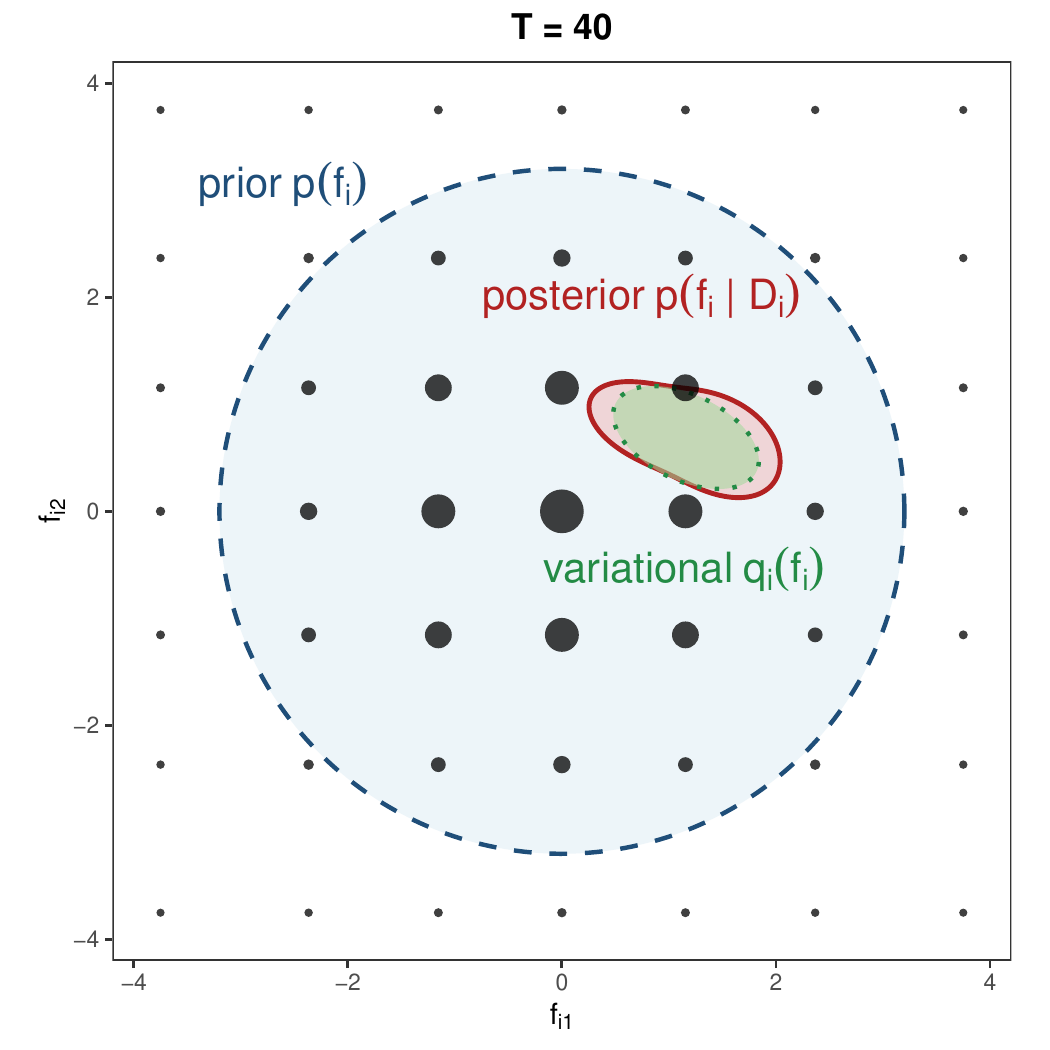}
    \caption{Schematic illustration of AVEM versus numerically implemented exact EM for a two-dimensional random effect \(f_i \in \mathbb{R}^2\). The large dashed circle represents a relatively diffuse prior distribution for \(f_i\), the solid red region represents the posterior distribution \(p(f_i \mid D_i)\), and the dotted green circle represents the variational approximation \(q_i(f_i)\). The black dots denote $7 \times 7$ Gaussian--Hermite nodes under the normal prior, with dot sizes proportional to the corresponding quadrature weights. As \(T_i\) increases, the posterior becomes more concentrated, so a fixed prior-centered Gaussian--Hermite grid becomes less efficient at resolving the dominant posterior region, whereas AVEM remains adapted to the posterior location and scale.}
    \label{fig:avem_exact_em_schematic}
\end{figure}

Figure~\ref{fig:avem_exact_em_schematic} provides a schematic illustration of why AVEM can be substantially more efficient than numerically implemented exact EM. In both panels, the posterior distribution is concentrated in a relatively small region compared with the prior, and this concentration becomes even more pronounced as \(T_i\) increases. Since Gaussian--Hermite quadrature uses a fixed set of prior-centered nodes, a large number of nodes may be required to accurately resolve the dominant posterior region when the posterior is sharply concentrated or shifted away from the prior mean. By contrast, AVEM directly constructs a variational approximation \(q_i(f_i)\) to the posterior itself, allowing it to adapt more effectively to the posterior location and scale.

From a computational perspective, AVEM only requires solving the forward--backward recursion once per subject at each iteration, whereas a numerically implemented exact EM algorithm must repeatedly evaluate the forward--backward recursion across many quadrature nodes or Monte Carlo samples. For example, Gaussian--Hermite quadrature requires \(J^d\) evaluations when \(J\) nodes are used in each dimension, while MCEM requires \(M\) evaluations for \(M\) Monte Carlo samples. Moreover, as discussed above, both \(J^d\) and \(M\) may need to be large when \(T_i\) is large and the posterior distribution becomes sharply concentrated. Therefore, although exact EM is theoretically less biased than a variational approximation, its practical numerical implementation may be computationally expensive and can itself suffer from substantial approximation error unless a sufficiently large number of nodes or samples is used.

\subsection{Connection to blocked mean-field variational inference} \label{sec:block_MFVI}

It is also natural to consider a blocked mean-field variational family \eqref{equ:block_MF},
which preserves the Markov dependence within \(U_i\) but breaks the posterior dependence between \(U_i\) and \(f_i\).
In the MHMM setting, the coordinate-wise optimal update for the latent state sequence satisfies
\begin{align*}
q_i^\ast(U_i)
&\propto
\exp\left\{
\E_{q_i(f_i)}
\left[
\log p(D_i,U_i,f_i;\theta)
\right]
\right\} \\
&\propto
\pi_{U_{i1}}
\left\{
\prod_{t=2}^{T_i}
\Gamma_{U_{i,t-1},U_{it}}
\right\}
\left\{
\prod_{t=1}^{T_i}
\exp\left(
\E_{q_i(f_i)}
\left[
\log e_{i,U_{it},t}(\theta,f_i)
\right]
\right)
\right\}.
\end{align*}
Equivalently, define the variational emission weights by
\[
\log \widetilde e_{ikt}(\theta)
=
\E_{q_i(f_i)}
\left[
\log e_{ikt}(\theta,f_i)
\right],
\qquad k=1,\ldots,K,\quad t=1,\ldots,T_i.
\]
Then 
\begin{align}
q_i^\ast(U_i)
\ \propto \ 
\pi_{U_{i1}}
\prod_{t=2}^{T_i}
\Gamma_{U_{i,t-1},U_{it}}
\prod_{t=1}^{T_i}
\widetilde e_{i,U_{it},t}(\theta).
\label{equ:qstar_form}
\end{align}
For comparison, conditional on a fixed value of \(f_i\), the latent path
posterior has the standard HMM form
\[
p(U_i\mid D_i,f_i;\theta)
\ \propto \ 
\pi_{U_{i1}}
\prod_{t=2}^{T_i}
\Gamma_{U_{i,t-1},U_{it}}
\prod_{t=1}^{T_i}
e_{i,U_{it},t}(\theta,f_i).
\]
Thus, \eqref{equ:qstar_form} preserves the HMM path structure, with the usual
emission likelihoods replaced by the variational emission weights
\(\widetilde e_{ikt}(\theta)\). Consequently, the variational state
probabilities and pairwise transition probabilities can be computed by the
standard forward--backward recursion.

This representation also clarifies the relation to AVEM. As detailed in
\eqref{equ:forward_update}--\eqref{equ:backward_update} of
Appendix~\ref{sec:forward_backward}, the forward--backward recursion for
\eqref{equ:qstar_form} is based on the variational emission weight
\[
\E_{q_i(f_i)}\!\left[\log e_{ikt}(\theta,f_i)\right].
\]
Mean-anchored AVEM instead uses the plug-in approximation
\[
\E_{q_i(f_i)}\!\left[\log e_{ikt}(\theta,f_i)\right]
\approx
\log e_{ikt}\!\left(\theta,\E_{q_i(f_i)}[f_i]\right).
\]
Thus, mean-anchored AVEM can be viewed as a plug-in approximation to the
blocked mean-field update for the latent state trajectory.

Compared with the anchored update, however, the blocked mean-field update is generally more computationally expensive. In each E-step, one must evaluate \(T_iK\) expectations with respect to \(q_i(f_i)\) for every subject, and in general these expectations require additional numerical integration or approximation. By contrast, AVEM only requires evaluating the emission terms at the anchor point and then running a single forward--backward recursion.

\subsection{Closed-form updates for Gaussian emissions}
\label{sec:special_case}

We next consider the Gaussian emission special case
\begin{align}
    D_{it}\mid (U_{it}=k,f_i) \sim N(\mu_k+f_i,\sigma_k^2 I_p),
\qquad
f_i\sim N(0,\Sigma). \label{equ:gauss_case}
\end{align}

The emission density is therefore
\begin{align*}
e_{ikt}(\theta,f_i)
&=
(2\pi\sigma_k^2)^{-p/2}
\exp\!\left\{
-\frac{1}{2\sigma_k^2}
\|D_{it}-\mu_k-f_i\|^2
\right\},
\end{align*}
where $\theta=(\pi,\Gamma,\{\mu_k,\sigma_k^2\}_{k=1}^K,\Sigma)$.

\paragraph{Variational E-step:}

For this Gaussian special case, the update of \(q_i(f_i)\) can be derived in closed form.
%
For subject $i$, the Gaussian variational distribution $q_i^{(m)}(f_i)$
has precision matrix
\begin{align*}
(\Omega_i^{(m)})^{-1}
&=
(\Sigma^{(m-1)})^{-1}
+
\sum_{t=1}^{T_i}\sum_{k=1}^K
\zeta_{ikt}^{(m)}(\sigma_k^{(m-1)})^{-2}I_p.
\end{align*}
and mean vector
\begin{align*}
\nu_i^{(m)}
&=
\Omega_i^{(m)}
\left[
\sum_{t=1}^{T_i}\sum_{k=1}^K
\zeta_{ikt}^{(m)} (\sigma_k^{(m-1)})^{-2}
\bigl(D_{it}-\mu_k^{(m-1)}\bigr)
\right].
\end{align*}

If the state-specific variances $\sigma_k^2$ are uniformly bounded above, then this precision matrix grows linearly with \(T_i\), and hence
\[
\tr(\Omega_i^{(m)})=O(T_i^{-1}).
\]
It then follows from Corollary~\ref{cor:mean_anchor_rate} that, in this special case, the suboptimality of using the mean-anchor is of order \(O(T_i^{-1})\).

\paragraph{Variational M-step}

Given \(\zeta_{ikt}^{(m)}\), \(\xi_{ik\ell t}^{(m)}\), and \(q_i^{(m)}(f_i)=N(\nu_i^{(m)},\Omega_i^{(m)})\), the M-step proceeds by updating the model parameters. As the updates for \(\pi\), \(\Gamma\), and \(\Sigma\) remain unchanged from the general case, we restrict attention here to the Gaussian emission parameters \(\mu_k\) and \(\sigma_k^2\) $(k = 1, \dots, K)$:
\begin{align*}
\mu_k^{(m)}
&=
\frac{
\sum_{i=1}^n\sum_{t=1}^{T_i}
\zeta_{ikt}^{(m)}
\bigl(D_{it}-\nu_i^{(m)}\bigr)
}{
\sum_{i=1}^n\sum_{t=1}^{T_i}
\zeta_{ikt}^{(m)}
},
\end{align*}

\begin{align*}
\sigma_k^{2\,(m)}
&=
\frac{
\sum_{i=1}^n\sum_{t=1}^{T_i}
\zeta_{ikt}^{(m)}
\left[
\|D_{it}-\mu_k^{(m)}-\nu_i^{(m)}\|^2
+
\tr(\Omega_i^{(m)})
\right]
}{
p\sum_{i=1}^n\sum_{t=1}^{T_i}\zeta_{ikt}^{(m)}
}.
\end{align*}

The above algorithm summarizes the proposed AVEM procedure. Details of the numerically implemented exact EM algorithm are provided in Appendix \ref{sec:mhmm_exact_em}.

\section{Anchored Variational EM for Mixed-effects State-space Models}
\label{sec:algorithm_ssm}

Next, we illustrate the proposed AVEM algorithm in another important special case, the mixed-effects state-space model (MESSM), see \cite{liu2011mixed, guo2024hierarchical}. 

Let $U_i=(U_{i1},\dots,U_{iT_i})$, $U_{it}\in\mathbb R^q$ denote the latent state process, and let
$D_i=(D_{i1},\dots,D_{iT_i})$, $D_{it}\in\mathbb R^p$ denote the observed process.
Subject \(i\) is assumed to follow the linear Gaussian state-space model 
\begin{align}
U_{i1} &\sim N(m_0,P_0), \notag\\
U_{it}\mid U_{i,t-1},f_i &\sim N(G_iU_{i,t-1},Q), \qquad t=2,\dots,T_i,
\label{eq:ressm_model_transition}\\
D_{it}\mid U_{it},f_i &\sim N(H_iU_{it},R), \qquad t=1,\dots,T_i, \notag
\end{align}
where \(G_i\in\mathbb R^{q\times q}\) is the subject-specific transition matrix, \(H_i\in\mathbb R^{p\times q}\) is the subject-specific loading matrix, and \(Q\in\mathbb R^{q\times q}\) and \(R\in\mathbb R^{p\times p}\) are the state and observation noise covariance matrices, respectively.

To ensure identifiability, we impose the normalization \(Q=I_q\) and constrain the loading matrix \(H_i\) to be lower triangular. These restrictions remove the usual rotational and scaling non-identifiability of the latent states, leaving only coordinate-wise sign changes in \(U_{it}\) \citep{wang2023latent}.
The sign non-identifiability may be addressed in two ways. First, one may use a careful initialization by fitting a reduced model with \(H_i = H_0\) and \(G_i = G_0\), and then using the resulting estimate of \(H_0\) and \(G_0\) to initialize the full model; this often substantially reduces sign switching. Second, one may enforce sign consistency across iterations by requiring each column of \(H_i\) to have positive cosine similarity with the corresponding column of the group-level loading matrix from the previous iteration.
See \citet{guo2024hierarchical} for further details. We further assume that \(R\) is diagonal.

To capture subject-specific heterogeneity, we introduce a Gaussian random effect 
\begin{align*}
    f_i =(g_i^{\top},h_i^{\top})^{\top}\sim N(\mu,\Sigma).
\end{align*}
where \(\mu=(\mu_g^{\top},\mu_h^{\top})^{\top}\), and \(\Sigma=\operatorname{diag}(\Sigma_g,\Sigma_h)\). Here \(g_i=\operatorname{vec}(G_i)\) denotes the vector obtained by stacking the columns of \(G_i\), and \(\mu_g=\operatorname{vec}(G)\) is the corresponding population-level mean. Likewise, \(h_i=\operatorname{vecl}(H_i)\) denotes the vector formed by stacking the free lower-triangular entries of \(H_i\), and \(\mu_h=\operatorname{vecl}(H)\) is the corresponding population-level mean. We use \(\operatorname{vec}(\cdot)\) for the usual column-wise vectorization of a matrix and \(\operatorname{vecl}(\cdot)\) for the vectorization of its free lower-triangular entries only. To relate these two representations for \(H_i\), let \(S_H\) denote the fixed expansion matrix such that
$\operatorname{vec}(H_i)=S_H h_i$.
Thus, the lower-triangular structure of \(H_i\) is enforced automatically through the reduced parameterization \(h_i=\operatorname{vecl}(H_i)\).

Under the random-effects state-space model described above, the complete-data joint log-likelihood for subject \(i\) is
\begin{align*}
\log p(D_i,U_i,f_i;\theta)
&=
-\frac12 \log |\Sigma|
-\frac12 (f_i-\mu)^\top \Sigma^{-1}(f_i-\mu)
 \\ 
&\quad -\frac12 \log |P_0|
-\frac12 (U_{i1}-m_0)^\top P_0^{-1}(U_{i1}-m_0)
\\
&\quad -\frac{T_i}{2}\log |R| -\frac12 \sum_{t=1}^{T_i}
(D_{it}-H_iU_{it})^\top R^{-1}(D_{it}-H_iU_{it})
\\
&\quad
-\frac12 \sum_{t=2}^{T_i}
(U_{it}-G_iU_{i,t-1})^\top (U_{it}-G_iU_{i,t-1})
+ \mathrm{const},
\end{align*}
where \(\theta=(m_0,P_0,R,\mu,\Sigma)\), \(Q=I_q\) is fixed for identifiability.

We adopt the anchored variational family
\[
q_i(U_i,f_i)=q_i(g_i) q_i(h_i)\,p(U_i\mid D_i,f_{0i};\theta),
\]
where \(f_{0i}\) is an anchor point. 

The corresponding anchored evidence lower bound for subject \(i\) is
\begin{align*}
\mathcal L_i^A(q_i,\theta;f_{0i})
&=
\mathbb E_{q_i(g_i)\,q_i(h_i)\,p(U_i\mid D_i,f_{0i};\theta)}
\Big[
\log p(D_i,U_i,f_i;\theta)
\nonumber\\
&\hspace{2.5cm}
-\log p(U_i\mid D_i,f_{0i};\theta)
-\log q_i(g_i)
-\log q_i(h_i)
\Big].
\end{align*}

\begin{algorithm}[ht]
\caption{AVEM algorithm for the RESSM}
\label{alg:avem_messm}
\begin{algorithmic}[1]
\Require Data $\{D_i\}_{i=1}^n$, initialize $r\gets 0$, $\theta^{(0)}$, $\{\nu_{g,i}^{(0)},\Omega_{g,i}^{(0)},\nu_{h,i}^{(0)},\Omega_{h,i}^{(0)}\}_{i=1}^n$

\State \textbf{E-step:}
\Repeat
    \For{$i=1,\dots,n$}
        \State Set $g_{0i}^{(r)}=\nu_{g,i}^{(r)}$, $h_{0i}^{(r)}=\nu_{h,i}^{(r)}$
        \State Update $p(U_i\mid D_i,f_{0i}^{(r)};\theta^{(r)})$
        and compute $\widehat m_{it}^{(r)}$, $\widehat P_{it}^{(r)}$, $\widehat P_{i,t,t-1}^{(r)}$, $\widehat Q_{it}^{(r)}$, and $\widehat Q_{i,t,t-1}^{(r)}$ using \eqref{eq:hatm_def}-\eqref{eq:ressm_cross_moment} via the Kalman filter and smoother

        \State Update $q_i(g_i)=N\!\left(\nu_{g,i}^{(r+1)},\Omega_{g,i}^{(r+1)}\right)$
        using \eqref{eq:Lambda_gi} and \eqref{eq:eta_gi}
        \State Update $q_i(h_i)=N\!\left(\nu_{h,i}^{(r+1)},\Omega_{h,i}^{(r+1)}\right)$
        using \eqref{eq:Lambda_hi} and \eqref{eq:eta_hi}
        \State (Optional) Align each column of \(H_i\) with the corresponding group-level column from the previous iteration, and adjust all associated quantities accordingly.
    \EndFor

    \State \textbf{M-step:} update $\theta = (\mu_g,\ \mu_h,\ \Sigma_g,\ \Sigma_h,\ m_0,\ P_0,\ R)$
    using \eqref{eq:mu_sigma_g_h}, \eqref{eq:m0_p0}, and \eqref{eq:Rj_update_final}
    \State $r\gets r+1$
\Until{relative ELBO convergence or maximum iterations reached}

\State \Return $\widehat\theta=\theta^{(r)}$
\end{algorithmic}
\end{algorithm}

We present the AVEM algorithm for MESSMs in Algorithm~\ref{alg:avem_messm}; the detailed derivation is deferred to Appendix~\ref{sec:app_avem_messm}. Notably, each step of the algorithm admits an explicit update, and only one Kalman filter/smoother pass is required for each EM iteration. This leads to substantial computational savings. In contrast, as noted by \citet{liu2011mixed}, the exact EM algorithm for MESSMs cannot in general be implemented through a fully efficient Kalman filter-based procedure, and numerical approximations such as MCEM are typically required to evaluate the conditional expectations. Another alternative is fully Bayesian inference, for example via Gibbs sampling, but such methods are often considerably slower when the primary goal is estimation or prediction rather than full posterior inference. 

Moreover, \citet{wang2004lack} showed that mean-field variational approximations may be inconsistent in this setting. Our AVEM approach therefore provides a useful middle ground, achieving both computational efficiency and good approximation accuracy.

\section{Simulation Studies}
\label{sec:simulation}

We conducted extensive simulation studies to evaluate the finite-sample performance of the proposed AVEM algorithm. We begin with a Gaussian special case, because it allows us to examine the behavior of the proposed method in a relatively simple setting where the emission model is continuous and easy to interpret. This experiment is designed to assess both statistical accuracy and computational efficiency under different sample sizes, sequence lengths, and levels of subject-specific heterogeneity.

\subsection{Scenario I: Gaussian MHMMs}

In Scenario I, we considered the Gaussian MHMM discussed in Section \ref{sec:special_case}. We assumed $f_i \sim N(0,\tau^2 I_d)$
and generated observations according to \eqref{equ:gauss_case}.
The latent state sequence was generated from a homogeneous first-order Markov chain with transition matrix $\Gamma$ and initial distribution $\delta$, taken to be the stationary distribution of $\Gamma$.

In the default setup, the state-specific mean vectors were specified as $\mu_k=a_k\mathbf{1}_d$, where $a_1,\dots,a_K$ are equally spaced values ranging from $1.5$ to $-1.5$. In particular, for $K=2$ we set $(a_1,a_2)=(1.5,-1.5)$, for $K=3$ we set $(a_1,a_2,a_3)=(1.5,0,-1.5)$, and for $K=4$ we set $(a_1,a_2,a_3,a_4)=(1.5,0.5,-0.5,-1.5)$. The state-specific variances were fixed at $\sigma_k^2=1$ for all $k$.
The transition matrix was chosen to be sticky, with diagonal entries equal to $0.92$ and off-diagonal entries equal to $0.08/(K-1)$.

\begin{figure}[ht]
    \centering
    \includegraphics[width=1\textwidth]{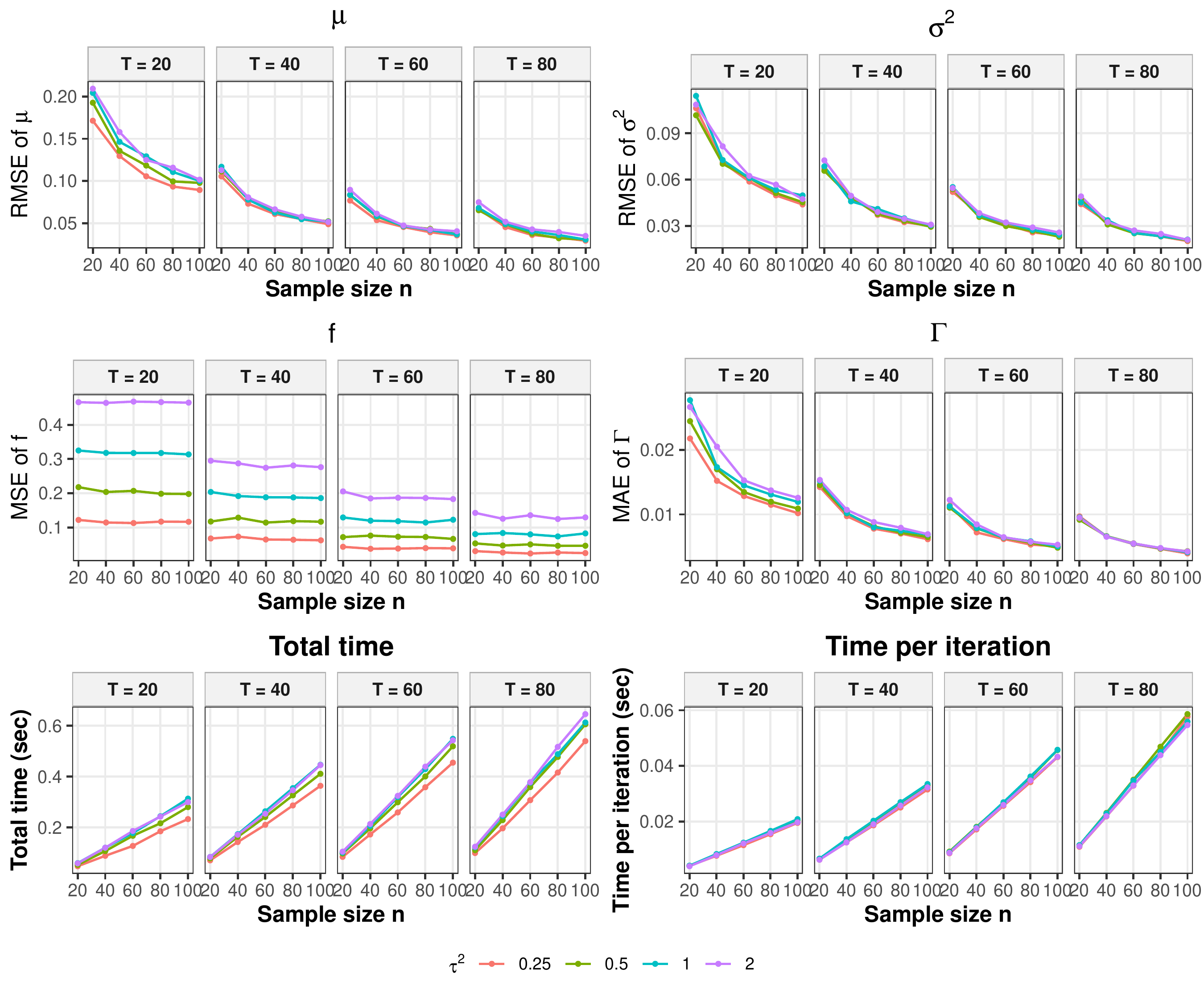}
    \caption{Estimation accuracy and computational performance of the proposed method under the Gaussian MHMM model with $K=3$ latent states and $d=2$, for varying sample sizes $n$, sequence lengths $T$, and random-effect variances $\tau^2$.}
    \label{fig:simu1_p1}
\end{figure}

In the first experiment, we studied estimation accuracy under varying sample size, sequence length, and random-effect variance. We fixed $K=3$ and $d=2$, and considered $n \in \{20,40,60,80,100\}$, $T \in \{20,40,60,80\}$, $\tau^2 \in \{0.25,0.5,1,2\}$.
For each combination of $(n,T,\tau^2)$, we generated $100$ Monte Carlo replicates.
We evaluated estimation accuracy using the RMSE for the state-dependent mean parameters $\mu_k$, the RMSE for the state-dependent variance parameters $\sigma_k^2$, and the average absolute entrywise error for the transition matrix $\Gamma$. To assess recovery of the subject-specific random effects, we also computed the mean squared error of $\widehat{f}_i$. 
All reported performance measures were averaged over the \(100\) replicates, and the results are summarized in Figure~\ref{fig:simu1_p1}.

Figure \ref{fig:simu1_p1} shows that estimation accuracy improves steadily as both the sample size $n$ and the sequence length $T$ increase. In particular, the estimation errors for $\mu_k$, $\sigma_k^2$, and $\Gamma$ decrease consistently with $n$, and are uniformly smaller for larger $T$. 
The random-effect variance $\tau^2$ has its strongest influence on the recovery of the subject-specific random effects $f_i$: larger values of $\tau^2$ lead to noticeably higher MSE for $\widehat f_i$, although the estimation error continues to decline as $T$ increases.
The computational time grows approximately linearly with both $n$ and $T$.

In the second experiment, we investigated how the estimation accuracy and computational cost of the proposed method vary with the number of latent states and the dimension of the subject-specific random effect. Specifically, we fixed the sample size $n=60$, sequence length $T=60$, and random-effect variance at $\tau^2=1$.
We varied the number of latent states over $K \in \{2,3,4,5,6,7\}$ and the dimension over $d \in \{2,3,4,5\}$.

\begin{figure}[ht]
    \centering
    \includegraphics[width=0.95\textwidth]{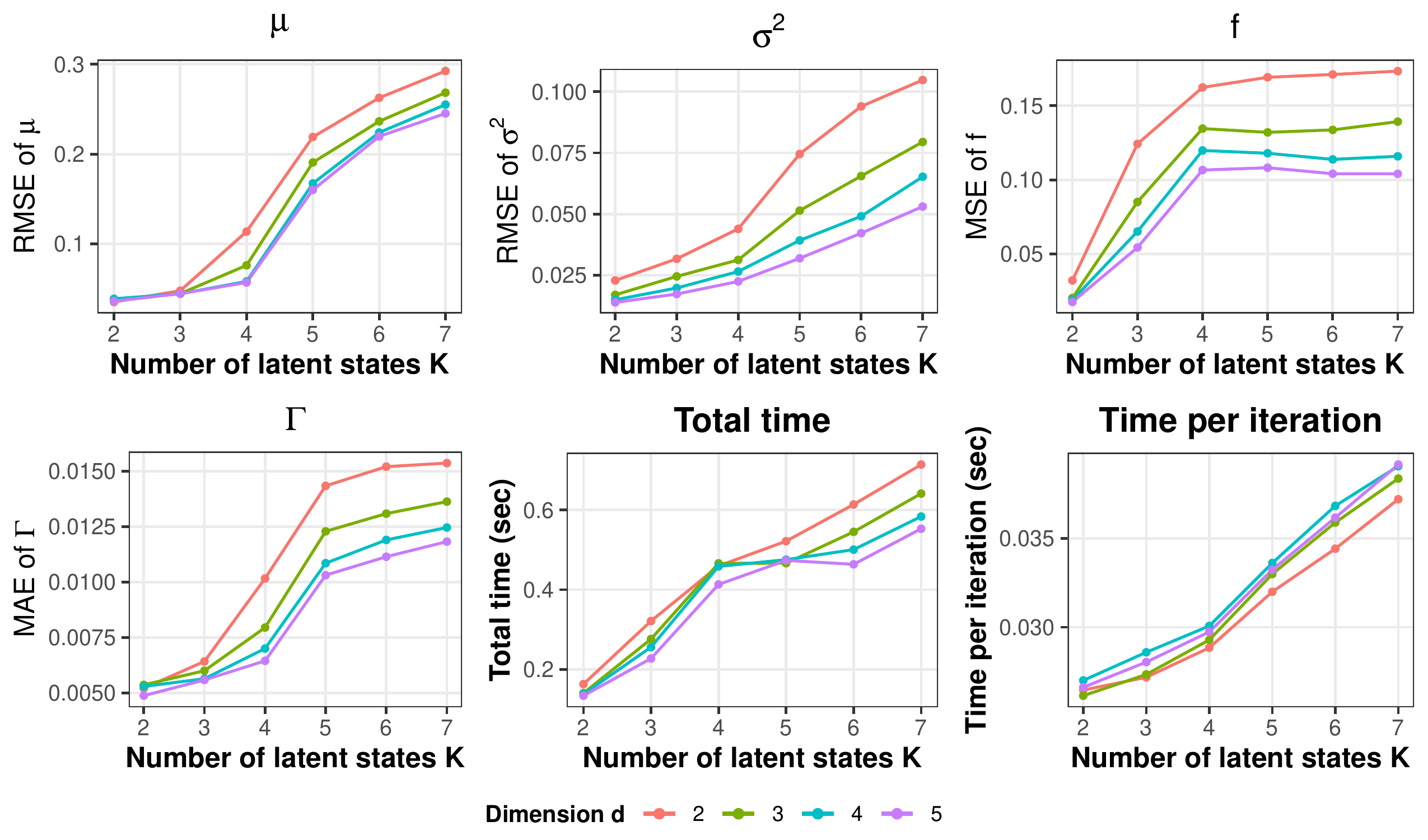}
    \caption{Estimation accuracy and computational performance of the proposed method under the Gaussian MHMM model with fixed $n=T=60$ and $\tau^2=1$, for varying numbers of latent states $K$ and random-effect dimensions $d$.}
    \label{fig:simu1_p2}
\end{figure}

Figure \ref{fig:simu1_p2} indicates that, for the Gaussian MHMM, runtime is driven mainly by the number of latent states $K$, rather than by the dimension $d$. Both the total computation time and the time per iteration increase steadily with $K$, while their dependence on $d$ remains relatively mild in this experiment. This is consistent with the structure of the Gaussian AVEM algorithm, where the dominant cost comes from the latent-state calculations, particularly the forward--backward step. Notably, even in the largest and most complex settings considered in both experiments, a single run of the proposed AVEM algorithm still takes less than one second on a 2023 MacBook Pro (Apple M3 Pro, single core), which underscores the practical computational advantage of the proposed approach.
This contrasts with the classical EM algorithm for MHMMs, where the E-step typically relies on either Gaussian quadrature or MCEM to integrate out the random effects.

To see this, we conducted a third experiment that compares the proposed AVEM algorithm with two standard classes of competitors: MCEM with different Monte Carlo sample sizes and quadrature-based EM (QEM) with different numbers of Gaussian--Hermite nodes. We considered settings with \(K=2\) latent states, \(n=40\) subjects, random-effect dimension \(d=2\), and variance parameter \(\tau^2 \in \{0.25, 1\} \), while varying the sequence length \(T \in \{20,40\}\). For MCEM, we used \(M\in\{25,50,100\}\) Monte Carlo samples, and for QEM, we used \(J\in\{3,5,7,9\}\) nodes per dimension. Across methods, we evaluated both estimation accuracy and computational efficiency.

\begin{figure}[ht]
    \centering
    \includegraphics[width=1.0\textwidth]{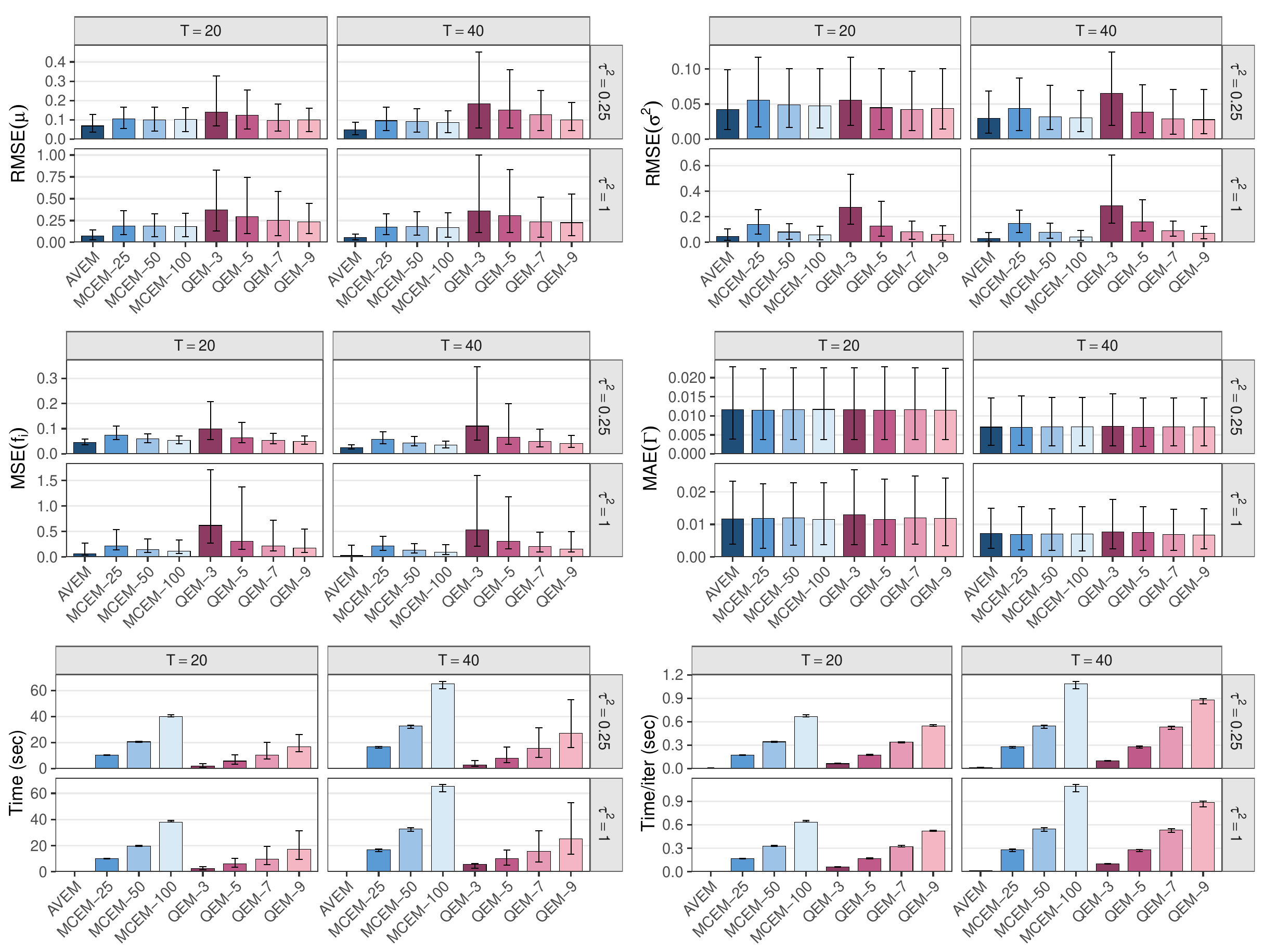}
    \caption{Performance comparison of AVEM, MCEM, and QEM for Gaussian MHMMs across different trajectory lengths $T$ and random-effect variances $\tau^2$. Bars show Monte Carlo medians, and error bars show empirical $90\%$ intervals over simulation replicates.}
    \label{fig:simu1_p3}
\end{figure}

Figure~\ref{fig:simu1_p3} demonstrates that AVEM provides the most favorable balance between estimation accuracy and computational efficiency.  Across all simulation settings, AVEM attains the smallest or nearly the smallest estimation errors, while requiring only a tiny fraction of the computation time needed by QEM and MCEM. Its advantage is particularly clear when \(\tau^2=1\), where AVEM yields substantially more accurate estimation of \(\mu\), \(\sigma^2\), and \(f_i\). When \(\tau^2=0.25\), the competing methods become closer in performance, but AVEM still remains the best overall in view of both accuracy and runtime. As expected, using more quadrature nodes or Monte Carlo samples generally improves the accuracy of QEM and MCEM, but only at a considerable computational cost. Taken together, these results show that AVEM is not only much faster, but also more accurate overall, especially in the more challenging scenarios with larger between-subject variability.


\subsection{Scenario II: Beyond Gaussian Emissions}
\label{subsec:sim_bern_mhmm}

To examine whether the anchored strategy remains effective beyond Gaussian emissions, we next consider a Bernoulli MHMM with a subject-specific random intercept. We take \(K=2\) latent states and use the same transition matrix and initial distribution as in Scenario~I.
For subject \(i\), we generate a scalar random effect $f_i \sim N(0,\tau^2)$
and then, conditional on \(U_{it}=k\) and \(f_i\), generate binary observations according to
\[
Y_{it} \mid (U_{it}=k,f_i)
\sim
\mathrm{Bernoulli}(p_{itk}),
\qquad
\logit(p_{itk})=\beta_k+f_i.
\]
We set \((\beta_1,\beta_2)=(-1.5,1.5)\), fix \(n=40\), and consider \(T\in\{100,200\}\) together with \(\tau^2\in\{0.25,1\}\). We compare AVEM with QEM using \(5\), \(10\), \(15\), and \(20\) Gaussian--Hermite nodes, and with MCEM using \(10\), \(15\), and \(20\) Monte Carlo samples. Since Scenario~I already illustrates the basic computational role of anchoring relative to numerical and Monte Carlo integration, the goal here is to assess whether the same qualitative pattern persists under a non-Gaussian emission model.

For this Bernoulli model, AVEM uses the Laplace approximation in \eqref{eq:laplace_q_update} to update the variational distribution of the subject-specific random effect, while the parameters \(\beta_k\) are updated by numerical optimization with Gaussian quadrature.

\begin{figure}[ht]
    \centering
    \includegraphics[width=1.0\textwidth]{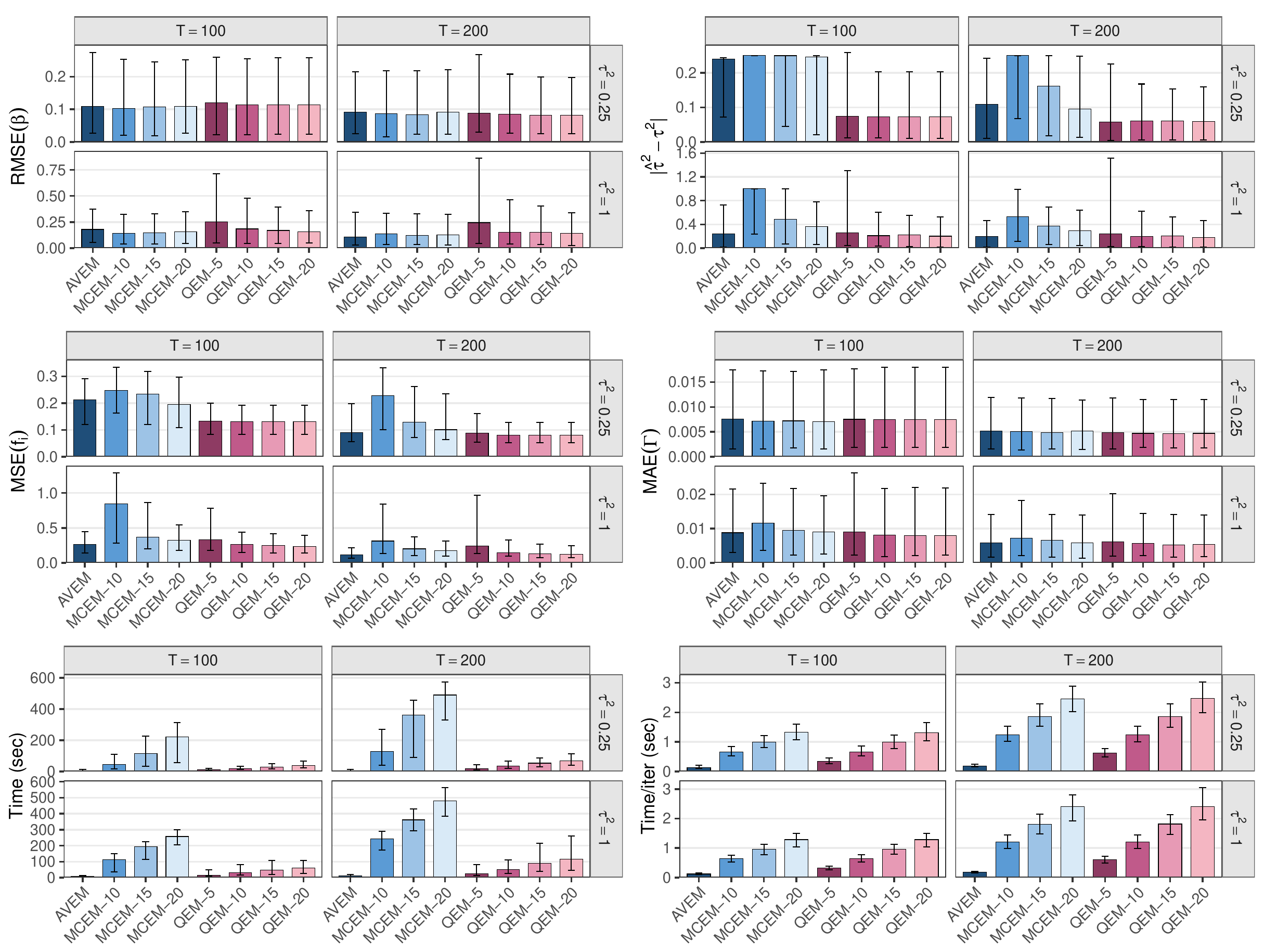}
    \caption{Performance comparison of AVEM, MCEM, and QEM for Bernoulli MHMMs across different trajectory lengths $T$ and random-effect variances $\tau^2$. Bars show Monte Carlo medians, and error bars show empirical $90\%$ intervals over simulation replicates.}
    \label{fig:bern_mhmm_sim}
\end{figure}

Figure~\ref{fig:bern_mhmm_sim} shows that the main computational advantage of AVEM remains clear in this Bernoulli setting. Across all four configurations, AVEM is substantially faster than both QEM and MCEM, while maintaining broadly competitive estimation accuracy. In particular, the advantage of AVEM is most pronounced in the more challenging settings with larger \(T\) and larger \(\tau^2\), where it often achieves the best overall balance between statistical accuracy and computational efficiency.

The main difference across methods appears in the estimation of \(\tau^2\). In several settings, QEM yields smaller estimation error for \(\tau^2\), especially when more quadrature nodes are used. By contrast, MCEM is noticeably less stable in the more challenging settings with \(\tau^2=1\), where Monte Carlo integration can occasionally lead to severe underestimation of the random-effect variance. Overall, these results indicate that the proposed AVEM algorithm continues to work well under a non-Gaussian emission model.

\subsection{Scenario III: MESSMs}
\label{subsec:sim_messm}

In Scenario III, we conduct a simulation study for the MESSM described in Section~\ref{sec:algorithm_ssm}. The data-generating mechanism follows \eqref{eq:ressm_model_transition}. 
We set \(q=2\) and \(p=4\), so that each subject has a two-dimensional latent state process and a four-dimensional observed process. The population mean transition and loading matrices are
\[
G=
\begin{pmatrix}
0.70 & -0.10\\
0.10 & 0.60
\end{pmatrix},
\qquad
H=
\begin{pmatrix}
1.0 & 0 \\
0.2 & 0.9 \\
0.3 & 0.4 \\
0.4 & 0.2
\end{pmatrix}.
\]
We further set \(U_{i1}\sim N(0,I_2)\) and \(R=0.25\,I_4\). Subject-specific matrices \(G_i\) and \(H_i\) are generated through Gaussian random effects on \(g_i=\operatorname{vec}(G_i)\) and \(h_i=\operatorname{vecl}(H_i)\), centered at \(G\) and \(H\), with covariance matrices \(\Sigma_g= 0.05 I_4\) and \(\Sigma_h=0.05 I_7\).

\begin{figure}[t]
    \centering
    \begin{subfigure}{0.95\textwidth}
        \centering
        \includegraphics[width=\textwidth]{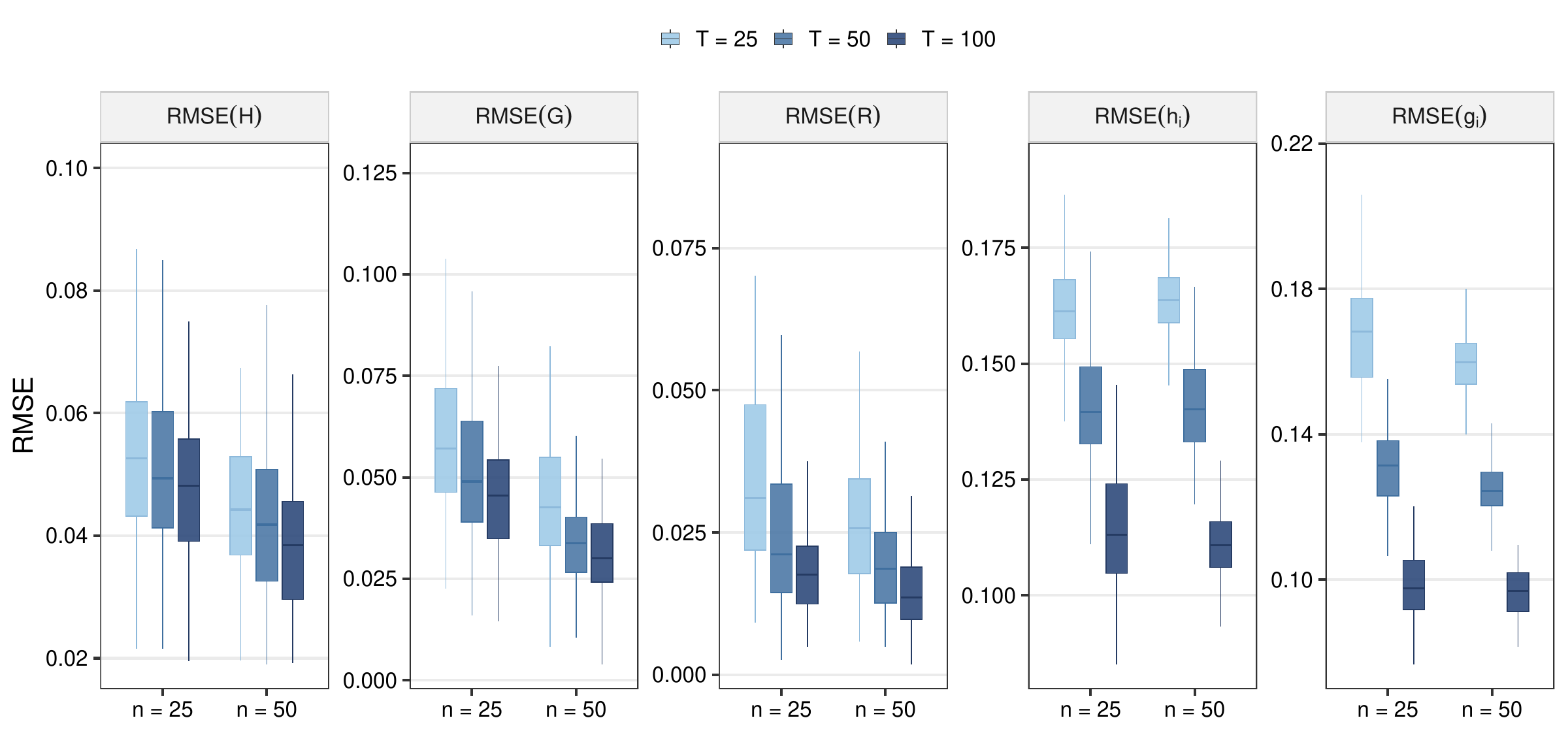}
    \end{subfigure}
    
    \vspace{0.8em}
    
    \begin{subfigure}{0.95\textwidth}
        \centering
        \includegraphics[width=\textwidth]{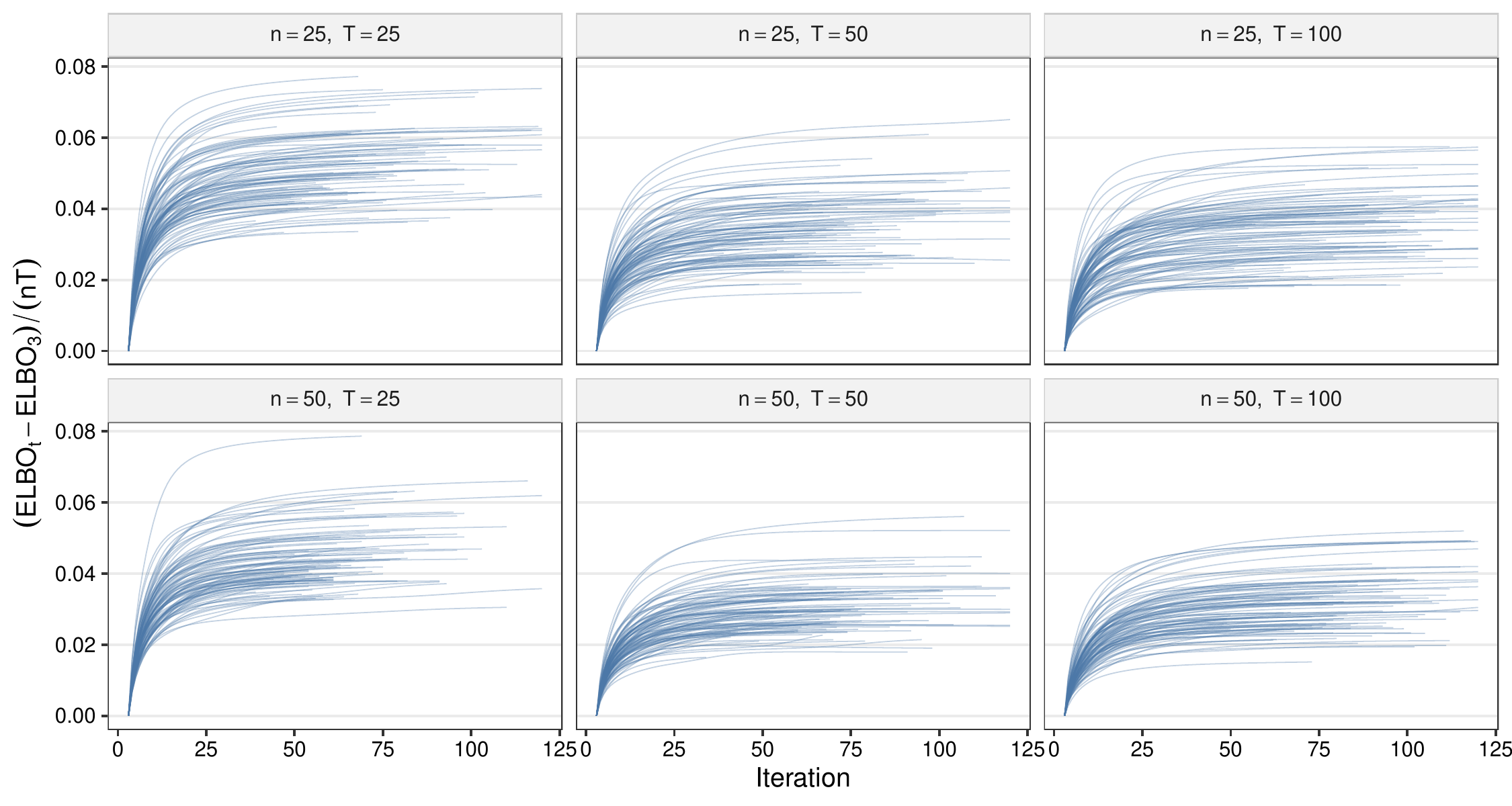}
    \end{subfigure}
    
    \caption{Top: boxplots of RMSEs under six settings. Bottom: normalized ELBO trajectories for the same six settings over \(100\) Monte Carlo replicates.}
    \label{fig:messm_sim}
\end{figure}

We consider six settings with \(n\in\{25,50\}\) and \(T\in\{25,50,100\}\), where $T_i \equiv T$. For each configuration, we generate \(100\) Monte Carlo data sets and fit the MESSM using Algorithm \ref{alg:avem_messm}. For each replicate, we record estimation errors for both population-level and subject-specific quantities. At the population level, we report the RMSE of the mean transition and loading matrices, together with the RMSE of the observation error covariance matrix \(R\). At the subject level, we report the RMSE of \(g_i\) and \(h_i\); for matrix-valued parameters, the RMSE is computed after vectorization. The boxplots in the top panel of Figure~\ref{fig:messm_sim} summarize these results. It shows that the proposed AVEM algorithm remains stable across all six settings and improves as both \(n\) and \(T\) increase.

In the bottom panel, we plot the normalized anchored ELBO, \((\mathrm{ELBO}_t-\mathrm{ELBO}_{t_0})/(nT)\), for each Monte Carlo replicate, with \(t_0=3\) to remove the instability in the first few iterations. The resulting trajectories are nearly monotone increasing across replicates, providing empirical support for the approximate monotonicity of the proposed AVEM updates and aligning well with the theory.

\section{Partially Anchored Variational Inference}
\label{subsec:partial_anchor}

The anchored variational inference framework is motivated by the idea that, for each subject \(i\), the posterior distribution of the subject-specific latent effect \(f_i\) becomes increasingly concentrated as the trajectory length \(T_i\) grows. In some applications, however, this concentration may hold only for part of \(f_i\), rather than for the entire vector. In such settings, a fully anchored approximation may be too restrictive, since it anchors components whose posteriors remain relatively diffuse.

To describe this situation, suppose that the subject-specific effect can be partitioned as
\[
f_i = \bigl(f_i^{(a)}, f_i^{(b)}\bigr),
\]
where \(f_i^{(a)}\) denotes a component that is well informed by the full trajectory and whose posterior becomes increasingly concentrated as \(T_i\) grows, while \(f_i^{(b)}\) denotes a component that remains more difficult to estimate. For example, this may occur when \(f_i^{(b)}\) influences only a short initial portion of the sequence, so that the amount of information about \(f_i^{(b)}\) does not continue to accumulate with \(T_i\). In such cases, the posterior distribution of \(f_i^{(b)}\) need not concentrate even as \(T_i \to \infty\).

This observation suggests a partially anchored extension of the variational approximation, in which only the component whose posterior is well concentrated is anchored. Specifically, one may consider a variational family of the form
\[
q_i(U_i,f_i^{(a)},f_i^{(b)})
=
q_i\bigl(f_i^{(a)}\bigr)\,
q_i\bigl(f_i^{(b)}\bigr)\,
p\!\left(U_i \mid D_i, f_{0i}^{(a)}, f_i^{(b)}\right),
\]
where \(f_{0i}^{(a)}\) is an anchor point for \(f_i^{(a)}\). Under this construction, the dependence on \(f_i^{(a)}\) is anchored, whereas the dependence on \(f_i^{(b)}\) is retained.

Conditional on the anchor value \(f_{0i}^{(a)}\), the posterior distribution of \(f_i^{(b)}\) is
\[
p(f_i^{(b)} \mid D_i, f_{0i}^{(a)})
=
\frac{p(D_i \mid f_{0i}^{(a)}, f_i^{(b)})\,p(f_i^{(b)})}
{\int p(D_i \mid f_{0i}^{(a)}, f_i^{(b)})\,p(f_i^{(b)}) \,df_i^{(b)}}.
\]
This naturally suggests approximating the variational factor \(q_i(f_i^{(b)})\) by the corresponding conditional posterior,
\[
q_i(f_i^{(b)})
=
p(f_i^{(b)} \mid D_i, f_{0i}^{(a)})
\ \propto \
p(D_i \mid f_{0i}^{(a)}, f_i^{(b)})\,p(f_i^{(b)}).
\]

Compared with the fully anchored approximation, the partially anchored construction retains flexibility in components of the latent effect whose posteriors remain diffuse. As a result, numerical integration or sampling is needed only over \(f_i^{(b)}\), rather than over the full vector \(f_i\), so the computational cost depends mainly on \(\dim(f_i^{(b)})\).

\begin{figure}[ht]
    \centering
    \begin{subfigure}[t]{0.47\textwidth}
        \centering
        \includegraphics[width=\textwidth]{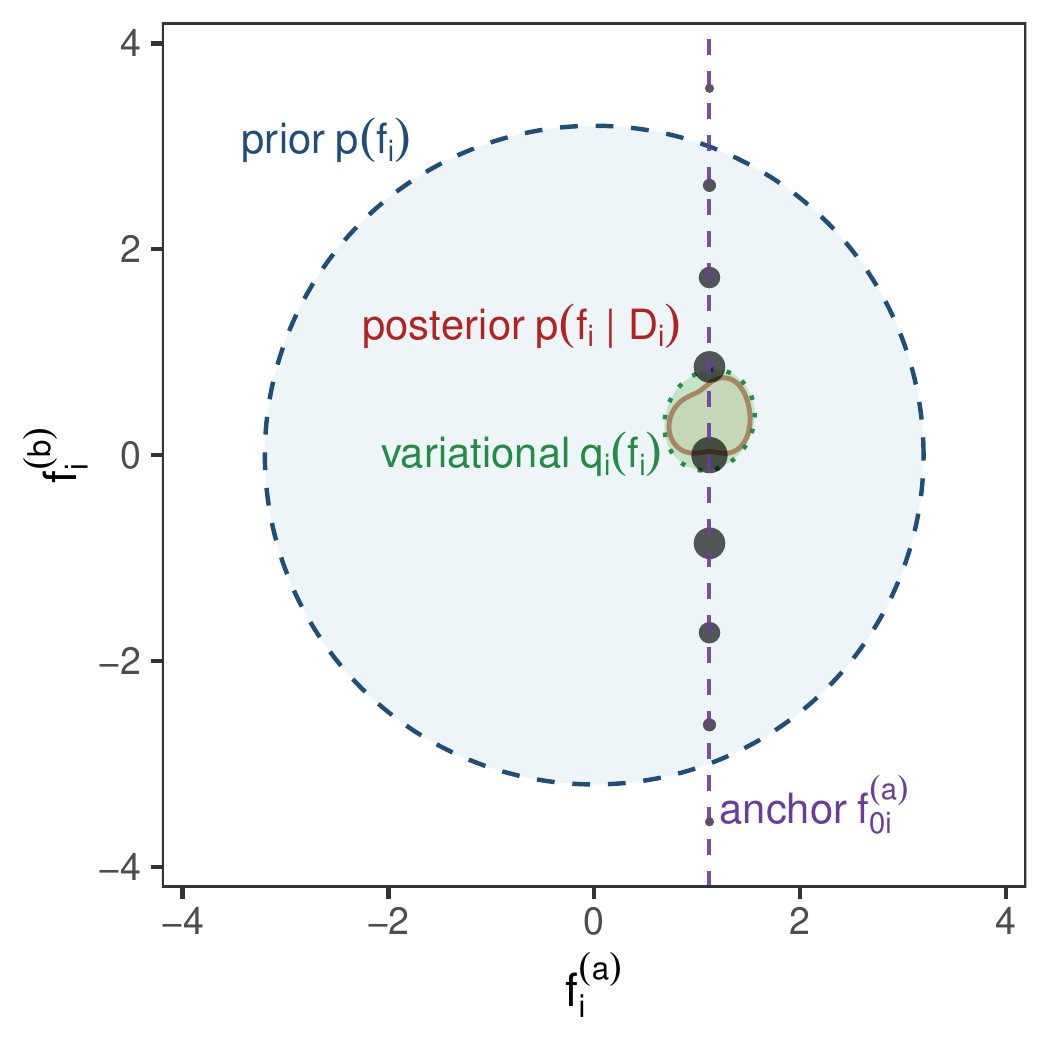}
        \caption{\(q_i(f_i^{(b)})\) concentrates.}
    \end{subfigure}\hfill
    \begin{subfigure}[t]{0.47\textwidth}
        \centering
        \includegraphics[width=\textwidth]{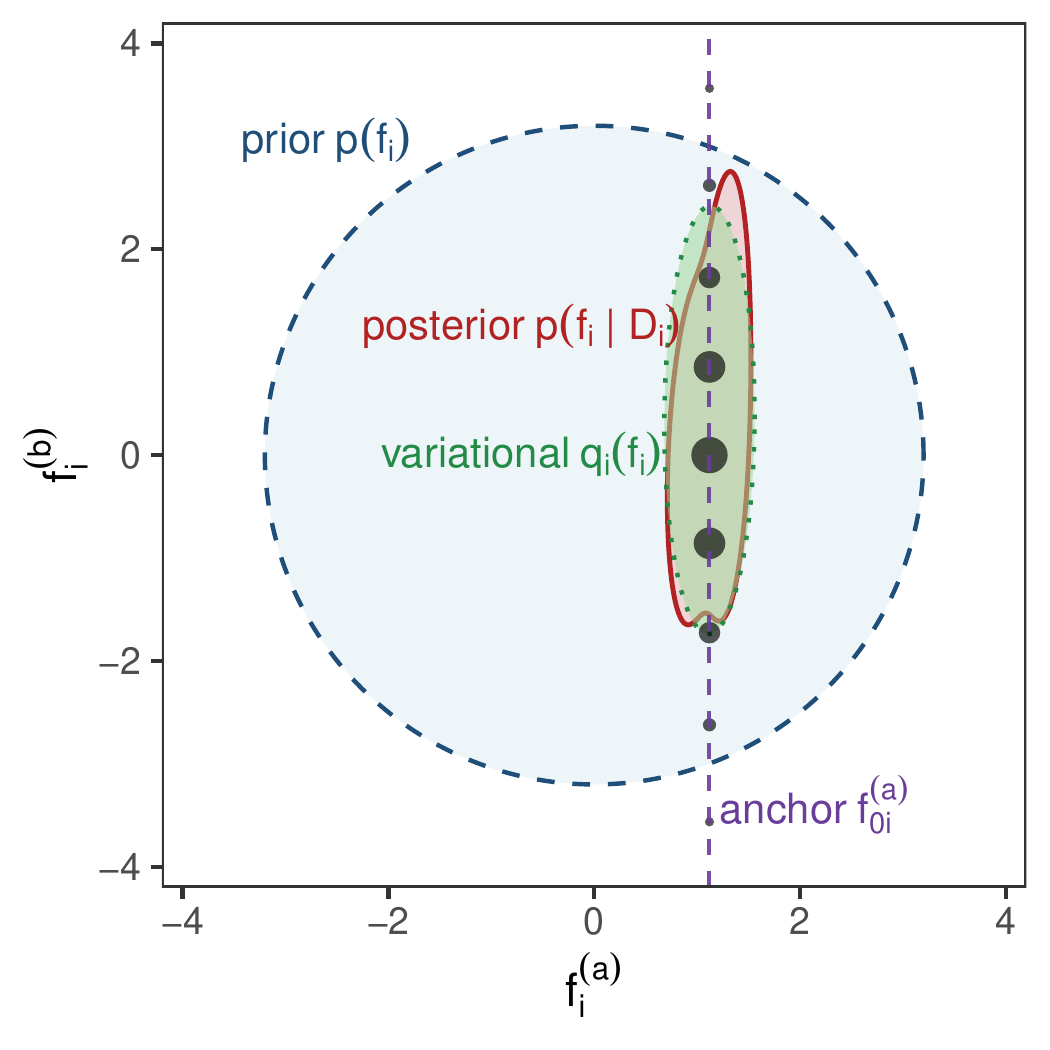}
        \caption{\(q_i(f_i^{(b)})\) does not concentrate.}
    \end{subfigure}
    \caption{Schematic illustration of partial anchoring for a two-dimensional random effect \(f_i = \bigl(f_i^{(a)}, f_i^{(b)}\bigr)\). The anchor point fixes \(f_{0i}^{(a)}\), while numerical integration is carried out only over \(f_i^{(b)}\).}
    \label{fig:partial_avem_schematic}
\end{figure}

Figure~\ref{fig:partial_avem_schematic} provides a schematic illustration. In both panels, the posterior is well concentrated in the \(f_i^{(a)}\) direction, so anchoring \(f_i^{(a)}\) at a representative value is reasonable. The difference lies in the behavior of \(f_i^{(b)}\): in the left panel, the posterior is also concentrated in that direction, whereas in the right panel it remains diffuse. The vertical line and the one-dimensional quadrature nodes indicate that partial anchoring fixes \(f_{0i}^{(a)}\) and performs numerical integration only over \(f_i^{(b)}\). Thus, when \(f_i^{(b)}\) does not become well concentrated, partial anchoring may offer a useful compromise between tractability and approximation accuracy.

We include this extension mainly to illustrate the flexibility of the anchored variational framework, while leaving a more detailed theoretical study of partially anchored inference to future work. A targeted simulation study for a localized-random-effect setting is reported in Appendix~\ref{app:sim_partial_anchor_localized}, where partial anchoring mainly improves estimation of the localized random effect, whereas the standard AVEM remains broadly competitive for the global parameters and is substantially faster.


\section{Discussion}
\label{sec:discussion}

In this paper, we proposed an anchored variational inference framework for personalized sequential latent-state models with subject-specific random effects. The main idea is to approximate the full conditional posterior of the local latent process by evaluating it at a representative value of the subject-specific latent effect, thereby preserving the tractable local structure of the original model while avoiding repeated recomputation over the full conditional family. This leads to a computationally efficient variational EM procedure that retains the essential dependence between the local latent trajectory and the subject-specific random effect, but replaces the full conditional dependence by an anchored approximation.

The proposed framework is particularly well suited to sequential settings, where each subject contributes a relatively long trajectory and the posterior distribution of the random effect tends to become increasingly concentrated as the sequence length grows. Under this concentration regime, the posterior mean provides a natural and theoretically justified anchor point. Our theoretical results show that the mean anchor is nearly optimal within the anchored family and that the resulting AVEM updates approximately preserve the monotonicity behavior of structured variational EM. These results help clarify why the anchored approximation is especially effective in longitudinal settings, where the subject-specific random effect is informed by many repeated observations from the same subject.

Using MHMMs and MESSMs as working examples, we showed that the proposed anchored variational framework leads to practical algorithms that still take advantage of the original computational structure of the model.
In the MHMM case, AVEM requires only a single forward--backward pass at the anchor point for each subject and iteration, rather than repeated evaluations over a large collection of quadrature nodes or Monte Carlo samples. In the MESSM case, the same principle yields an efficient procedure built on Kalman filtering and smoothing. The simulation results indicate that this reduction in computational burden can be substantial, while still maintaining good estimation accuracy. 
In particular, the gains are largest when the posterior distribution of the random effect is highly concentrated, since repeated numerical integration is least efficient in this setting.

At the same time, the proposed framework has several limitations. First, the quality of the anchored approximation depends on the extent to which the posterior distribution of the random effect is sufficiently concentrated. When the within-subject sequence is short, when the random effect is only weakly identified, or when the posterior is highly skewed or multimodal, a single anchor point may no longer provide an adequate summary of the relevant conditional dependence structure.
This also motivate the partially anchored extension introduced in Section~\ref{subsec:partial_anchor}. 
Second, although anchoring greatly reduces the computational burden, the update of the variational distribution in the E-step and the parameter update in the M-step may still not have closed-form solutions. Consequently, one may still need Monte Carlo approximation, numerical integration, or numerical optimization to carry out these steps in more complex models.

Several directions for future work merit further study. On the theoretical side, an important next step is to identify conditions under which the AVEM objective and the corresponding structured variational objective have the same, or asymptotically equivalent, local optima. 
On the methodological side, it would be useful to develop practical diagnostics for determining when anchoring is appropriate and when it should be avoided. Finally, although the present paper has focused on MHMMs and MESSMs, the same idea may also extend beyond sequential settings, such as multi-site studies and other hierarchical models with tractable conditional local inference and low-dimensional subject-specific heterogeneity. We hope that the proposed framework provides a useful step toward scalable and principled inference in this broader class of models.

\acks{The R implementations of the AVEM algorithms for MHMMs and MESSMs are available in the GitHub repository at \url{https://github.com/xingcheg/AVEM}}

\appendix

\section*{Appendix}

\section{More Details on the MHMMs} 

\subsection{Forward-backward recursions for efficient HMM posterior computation} \label{sec:forward_backward}

Denote
\[
\zeta_{ikt}(\theta,f_i)=p(U_{it}=k\mid D_i,f_i;\theta),
\qquad
\xi_{ik\ell t}(\theta,f_i)=p(U_{it}=k,U_{i,t+1}=\ell\mid D_i,f_i;\theta).
\]
Recall that \(e_{ikt}(\theta,f_i)\), defined in \eqref{equ:emission}, denotes the emission density at trial \(t\) under latent state \(k\).
For notational simplicity, in the remainder of this subsection we suppress the explicit dependence of \(\zeta_{ikt}\), \(\xi_{ik\ell t}\), and \(e_{ikt}\) on \(\theta\) and \(f_i\).

Recall that the latent state process has initial distribution \(\pi_k\) and transition matrix \(\Gamma=(\Gamma_{k\ell})\).

Define the forward variable: $a_{ikt} = p(U_{it}=k, D_{i[1:t]}\mid f_i; \theta)$,
then
\begin{align}
    a_{ik1} = \pi_k e_{ik1} \qquad \mbox{and} \qquad a_{ikt}
=
e_{ikt}
\sum_{\ell=1}^K
a_{i \ell,t-1}\Gamma_{\ell k} \quad \mbox{for} \quad t=2,\dots,T_i. \label{equ:forward_update}
\end{align}

Similarly, define the backward variable
$
b_{ikt} = p(D_{i[t+1:T_i]}\mid U_{it}=k,f_i,D_{i[1:t]}; \theta),$
then
\begin{align}
   b_{ikT_i}=1 \qquad \mbox{and} \qquad b_{ikt}
=
\sum_{\ell=1}^K
\Gamma_{k\ell} e_{i\ell,t+1} b_{i\ell,t+1} \quad \mbox{for} \quad t=T_i-1,\dots,1. \label{equ:backward_update} 
\end{align}

Hence the posterior state probabilities are
\begin{align}
    \zeta_{ikt}
=
\frac{a_{ikt} b_{ikt}}{\sum_{j=1}^K a_{ijT_i}}
\qquad \mbox{and} \qquad
\xi_{i k \ell t}
=
\frac{
a_{ikt}\Gamma_{k\ell}e_{i \ell,t+1} b_{i \ell,t+1}
}{
\sum_{j=1}^K a_{ijT_i}
}. \label{equ:hmm_posterior}
\end{align}

As a by-product of the forward--backward algorithm, we obtain the conditional
marginal likelihood
\[
p(D_i \mid f_i,\theta)
=
\sum_{k=1}^K p(U_{iT_i}=k,D_{i} \mid f_i,\theta)
=
\sum_{k=1}^K a_{ikT_i}.
\]

In practice, the forward--backward recursions involve repeated multiplication of small probabilities, which can easily lead to numerical underflow. To improve numerical stability, one may work on the log scale, where products become sums and divisions become differences, and then exponentiate back after normalization.

\subsection{Numerically implemented exact EM for Gaussian MHMMs} \label{sec:mhmm_exact_em}

For the numerical implementation of the EM algorithm, we assume without loss of generality that \(T_i = T\) for all \(i\). In addition, we take \(p=d\), so that the observed vector \(D_{it}\) and the random effect \(f_i\) have the same dimension. We further simplify that \(\Sigma=\tau^2 I_d\).

Then the EM objective at iteration \(m\) is
\begin{align*}
Q(\theta \mid \theta^{(m-1)})
&=
\sum_{i=1}^n
\sum_{t=1}^{T}\sum_{k=1}^K
\E_{p(f_i \mid D_i; \theta^{(m-1)})}
\Big[ \zeta_{ikt}(\theta^{(m-1)}, f_i) \log e_{ikt}(\theta,f_i) \Big] \\
&\quad
+
\sum_{i=1}^n \sum_{k=1}^K
\E_{p(f_i \mid D_i; \theta^{(m-1)})}
\Big[\zeta_{ik1}(\theta^{(m-1)}, f_i)\Big]\log \pi_k \\
&\quad
+
\sum_{i=1}^n \sum_{t=1}^{T-1}\sum_{k=1}^K\sum_{\ell=1}^K
\E_{p(f_i \mid D_i; \theta^{(m-1)})}
\Big[\xi_{ik\ell t}(\theta^{(m-1)}, f_i)\Big]\log \Gamma_{k\ell} \\
&\quad
+
\sum_{i=1}^n
\E_{p(f_i \mid D_i; \theta^{(m-1)})}
\Big[\log p(f_i;\tau^2)\Big].
\end{align*}

Moreover,
\begin{align*}
p(f_i \mid D_i; \theta)
&=
\frac{p(D_i \mid f_i; \theta)\, p(f_i; \theta)}
{\int p(D_i \mid f; \theta)\, p(f; \theta)\, df},
\quad \mbox{where} \quad
p(D_i \mid f; \theta)=\sum_{j=1}^K a_{ijT}(\theta,f).
\end{align*}

Numerically, let \(\{f_j\}_{j=1}^J\) denote either a set of quadrature nodes or a set of Monte Carlo samples. We approximate the posterior distribution of \(f_i\) by a discrete distribution supported on \(\{f_j\}_{j=1}^J\), with weights
\[
\hat{w}_{ij}^{(m)}
\ \propto \ 
v_j\, p\!\left(D_i \mid f_j; \theta^{(m-1)}\right),
\qquad j=1,\dots,J,
\]
and normalize them so that \(\sum_{j=1}^J \hat{w}_{ij}^{(m)}=1\). Here \(v_j\) denotes the numerical weight associated with node \(f_j\): in MCEM, \(v_j \equiv 1\), whereas in QEM, \(v_j\) is the corresponding Gaussian--Hermite quadrature weight.

Using these weights, we compute
\begin{align*}
\bar{\zeta}_{ikt}^{(m)}
&=
\sum_{j=1}^J \zeta_{ikt}\!\left(\theta^{(m-1)}, f_j\right)\, \hat{w}_{ij}^{(m)},
\qquad
\bar{\xi}_{ik\ell t}^{(m)}
=
\sum_{j=1}^J \xi_{ik\ell t}\!\left(\theta^{(m-1)}, f_j\right)\, \hat{w}_{ij}^{(m)}.
\end{align*}

The updates for the initial and transition probabilities are
\begin{align*}
\pi_k^{(m)}
&=
\frac{1}{n}\sum_{i=1}^n \bar{\zeta}_{ik1}^{(m)},
\qquad
\Gamma_{k\ell}^{(m)}
=
\frac{
\sum_{i=1}^n \sum_{t=1}^{T-1} \bar{\xi}_{ik\ell t}^{(m)}
}{
\sum_{i=1}^n \sum_{t=1}^{T-1} \bar{\zeta}_{ikt}^{(m)}
}.
\end{align*}

The updates for the Gaussian emission parameters are
\begin{align*}
\mu_k^{(m)}
&=
\frac{
\sum_{j=1}^J \sum_{i=1}^n \sum_{t=1}^{T}
\hat{w}_{ij}^{(m)} \zeta_{ikt}\!\left(\theta^{(m-1)}, f_j\right)\, \bigl(D_{it} - f_j\bigr)
}{
\sum_{j=1}^J \sum_{i=1}^n \sum_{t=1}^{T}
\hat{w}_{ij}^{(m)} \zeta_{ikt}\!\left(\theta^{(m-1)}, f_j\right)
},
\\[1em]
\sigma_k^{2\,(m)}
&=
\frac{
\sum_{j=1}^J \sum_{i=1}^n \sum_{t=1}^{T}
\hat{w}_{ij}^{(m)} \zeta_{ikt}\!\left(\theta^{(m-1)}, f_j\right)\,
\|D_{it}-\mu_k^{(m)}-f_j\|^2
}{
d \sum_{j=1}^J \sum_{i=1}^n \sum_{t=1}^{T}
\hat{w}_{ij}^{(m)} \zeta_{ikt}\!\left(\theta^{(m-1)}, f_j\right)
}.
\end{align*}

Finally, since \(\Sigma=\tau^2 I_d\), the update for \(\tau^2\) is
\begin{align*}
\tau^{2(m)}
&=
\frac{1}{nd}
\sum_{i=1}^n
\sum_{j=1}^J
\hat{w}_{ij}^{(m)}\, f_j^{\top} f_j.
\end{align*}

\section{AVEM algorithm for the MESSMs} \label{sec:app_avem_messm}

\paragraph{Kalman filter and smoother under the anchor $f_{0i}$}
Given \(\theta\) and the anchor \(f_{0i}=(g_{0i}^\top,h_{0i}^\top)^\top\), the conditional model for \(U_i=(U_{i1}^\top,\dots,U_{iT_i}^\top)^\top\) is linear Gaussian. Hence the conditional posterior $p(U_i\mid D_i,f_{0i};\theta)$
can be computed by the Kalman filter and smoother.
Let
\[
m_{it}^{\,s}:=\mathbb E(U_{it}\mid D_{i[1:s]},f_{0i};\theta),
\qquad
P_{it}^{\,s}:=\Var(U_{it}\mid D_{i[1:s]},f_{0i};\theta),
\]
where \(D_{i[1:s]}\) denotes the observations for subject \(i\) up to time \(s\). In particular,
\(m_{it}^{\,t-1}\) and \(P_{it}^{\,t-1}\) are the one-step-ahead predicted mean and covariance, while
\(m_{it}^{\,t}\) and \(P_{it}^{\,t}\) are the filtered mean and covariance.

The Kalman filter recursions are
\begin{align*}
&m_{it}^{\,t-1}
= G_i m_{i,t-1}^{\,t-1}, \quad
P_{it}^{\,t-1}
= G_i P_{i,t-1}^{\,t-1} G_i^\top+I_q,\\
&K_{it}
= P_{it}^{\,t-1}H_{i}^\top
\Bigl(H_{i}P_{it}^{\,t-1}H_{i}^\top+R\Bigr)^{-1},\\
&m_{it}^{\,t}
= m_{it}^{\,t-1}
+K_{it}\Bigl(D_{it}-H_{i}m_{it}^{\,t-1}\Bigr),
\\
&P_{it}^{\,t}
= \bigl(I-K_{it}H_{i}\bigr)P_{it}^{\,t-1}.
\end{align*}

After the forward filtering pass, the backward smoother is initialized at \(t=T_i\)
and for \(t=T_i,\dots,2\), the recursions are
\begin{align*}
J_{i,t-1}
&= P_{i,t-1}^{\,t-1}G_i^\top\bigl(P_{it}^{\,t-1}\bigr)^{-1},\\
m_{i,t-1}^{\,T_i}
&= m_{i,t-1}^{\,t-1}
+J_{i,t-1}\bigl(m_{it}^{\,T_i}-m_{it}^{\,t-1}\bigr),\\
P_{i,t-1}^{\,T_i}
&= P_{i,t-1}^{\,t-1}
+J_{i,t-1}\bigl(P_{it}^{\,T_i}-P_{it}^{\,t-1}\bigr)J_{i,t-1}^\top.
\end{align*}

Therefore, the posterior moments appearing in the anchored local update are
\begin{align}
\widehat m_{it}
&= m_{it}^{\,T_i}
= \mathbb E(U_{it}\mid D_i,f_{0i};\theta),
\label{eq:hatm_def}\\
\widehat P_{it}
&= P_{it}^{\,T_i}
= \Var(U_{it}\mid D_i,f_{0i};\theta).
\label{eq:hatP_def}
\end{align}
In addition, the lag-one smoothed cross-covariance is
\begin{align}
\widehat P_{i,t,t-1}
&=
\Cov(U_{it},U_{i,t-1}\mid D_i,f_{0i};\theta) =
\widehat P_{it} J_{i,t-1}^\top.
\label{eq:rts_lagone_cov}
\end{align}
Consequently,
\begin{align}
\mathbb E_0(U_{it}U_{it}^\top)
&:= \widehat Q_{it} = 
\widehat P_{it}+\widehat m_{it}\widehat m_{it}^\top,
\label{eq:ressm_second_moment}\\
\mathbb E_0(U_{it}U_{i,t-1}^\top)
&:= \widehat Q_{i,t,t-1} =
\widehat P_{i,t,t-1}+\widehat m_{it}\widehat m_{i,t-1}^\top,
\label{eq:ressm_cross_moment}
\end{align}
where \(\mathbb E_0(\cdot)\) denotes expectation under \(p(U_i\mid D_i,f_{0i};\theta)\).


\paragraph{Variational update for \(q_i(g_i)\).}

The optimal update for \(q_i(g_i)\) satisfies
\begin{align*}
\log q_i^*(g_i)
&=
\mathbb E_{q_i(h_i)\,p(U_i\mid D_i,f_{0i};\theta)}
\!\left[\log p(D_i,U_i,f_i;\theta)\right]
+\mathrm{const}
\\
&=
-\frac12 (g_i-\mu_g)^\top \Sigma_g^{-1}(g_i-\mu_g)
-\frac12 \sum_{t=2}^{T_i}
\mathbb E_0
\Big[
\bigl(U_{it}-B_{it}^{(G)}g_i\bigr)^\top
\bigl(U_{it}-B_{it}^{(G)}g_i\bigr)
\Big]
+\mathrm{const},
\end{align*}
where \(B_{it}^{(G)}=U_{i,t-1}^\top \otimes I_q\), so that \(G_iU_{i,t-1}=B_{it}^{(G)}g_i\).
Hence 
\begin{align*}
    q_i^*(g_i)=N( \nu_{g,i},\, \Omega_{g,i}) = N(\Lambda_{g,i}^{-1} \eta_{g,i},\, \Lambda_{g,i}^{-1}),
\end{align*}
where
\begin{align}
\Lambda_{g,i}
&=
\Sigma_g^{-1}
+
\sum_{t=2}^{T_i}
\mathbb E_0\!\left[(B_{it}^{(G)})^\top B_{it}^{(G)}\right]
\nonumber\\
&=
\Sigma_g^{-1}
+
\sum_{t=2}^{T_i}
\widehat Q_{i,t-1} \otimes I_q,
\label{eq:Lambda_gi}\\
\eta_{g,i}
&=
\Sigma_g^{-1}\mu_g
+
\sum_{t=2}^{T_i}
\mathbb E_0\!\left[(B_{it}^{(G)})^\top U_{it}\right]
\nonumber\\
&=
\Sigma_g^{-1}\mu_g
+
\sum_{t=2}^{T_i}
\operatorname{vec}\!\left( \widehat Q_{i,t,t-1} \right).
\label{eq:eta_gi}
\end{align}

\paragraph{Variational update for \(q_i(h_i)\).}
The optimal update for \(q_i(h_i)\) satisfies
\begin{align*}
\log q_i^*(h_i)
&=
\mathbb E_{p(U_i\mid D_i,f_{0i};\theta)}
\!\left[\log p(D_i,U_i,f_i;\theta)\right]
+\mathrm{const}\\
&=
-\frac12 (h_i-\mu_h)^\top \Sigma_h^{-1}(h_i-\mu_h)
\nonumber \\
&\quad\quad-\frac12 \sum_{t=1}^{T_i}
\mathbb E_0
\Big[
\bigl(D_{it}-B_{it}^{(H)}h_i\bigr)^\top
R^{-1}
\bigl(D_{it}-B_{it}^{(H)}h_i\bigr)
\Big]
+\mathrm{const},
\end{align*}
where $B_{it}^{(H)}=(U_{it}^\top \otimes I_p)S_H$, so that 
\begin{align*}
H_iU_{it}
=
(U_{it}^\top \otimes I_p)\operatorname{vec}(H_i)
=
(U_{it}^\top \otimes I_p)S_H h_i = B_{it}^{(H)}h_i.
\end{align*}

Hence 
\begin{align*}
    q_i^*(h_i)=N( \nu_{h,i},\, \Omega_{h,i}) = N(\Lambda_{h,i}^{-1} \eta_{h,i},\, \Lambda_{h,i}^{-1}),
\end{align*}
where

\begin{align}
\Lambda_{h,i}
&=
\Sigma_h^{-1}
+
\sum_{t=1}^{T_i}
\mathbb E_0\!\left[(B_{it}^{(H)})^\top R^{-1} B_{it}^{(H)}\right]
\nonumber \\
&=
\Sigma_h^{-1}
+
S_H^{\top} \left\{ \sum_{t=1}^{T_i}
\widehat Q_{it}  \otimes R^{-1} \right\} S_H,
\label{eq:Lambda_hi}\\
\eta_{h,i}
&=
\Sigma_h^{-1}\mu_h
+
\sum_{t=1}^{T_i}
\mathbb E_0\!\left[(B_{it}^{(H)})^\top R^{-1} D_{it}\right]
\nonumber \\
&=
\Sigma_h^{-1}\mu_h
+
S_H^{\top} \sum_{t=1}^{T_i}
\widehat m_{it} \otimes (R^{-1} D_{it}) .
\label{eq:eta_hi}
\end{align}

After updating \(q_i(g_i)\) and \(q_i(h_i)\), we refresh the anchor by the mean-anchor rule.

\paragraph{M-step:}
Given the anchored local posterior and the Gaussian variational factors \(q_i(g_i)\) and \(q_i(h_i)\), the M-step updates \(\theta\) by maximizing
\begin{align*}
\theta^{\mathrm{new}}
=
\arg\max_{\theta}
\sum_{i=1}^n
\mathbb E_{q_i(g_i)\,q_i(h_i)\,p(U_i\mid D_i,f_{0i};\theta^{\mathrm{old}})}
\bigl[
\log p(D_i,U_i,f_i;\theta)
\bigr].
\end{align*}
The updates for the prior parameters are
\begin{align}
\mu_g^{\mathrm{new}}
&=
\frac1n\sum_{i=1}^n \nu_{g,i},
\quad
\mu_h^{\mathrm{new}}
=
\frac1n\sum_{i=1}^n \nu_{h,i},
\nonumber
\\
\Sigma_g^{\mathrm{new}}
&=
\frac1n\sum_{i=1}^n
\left\{
\Omega_{g,i}+(\nu_{g,i}-\mu_g^{\mathrm{new}})(\nu_{g,i}-\mu_g^{\mathrm{new}})^\top
\right\},
\label{eq:mu_sigma_g_h}
\\
\Sigma_h^{\mathrm{new}}
&=
\frac1n\sum_{i=1}^n
\left\{
\Omega_{h,i}+(\nu_{h,i}-\mu_h^{\mathrm{new}})(\nu_{h,i}-\mu_h^{\mathrm{new}})^\top
\right\}.\nonumber
\end{align}
The updates for the initial-state parameters are
\begin{align}
m_0^{\mathrm{new}}
&=
\frac1n\sum_{i=1}^n \widehat m_{i1}, \quad
P_0^{\mathrm{new}}
=
\frac1n\sum_{i=1}^n
\Bigl\{
\widehat P_{i1}+(\widehat m_{i1}-m_0^{\mathrm{new}})(\widehat m_{i1}-m_0^{\mathrm{new}})^\top
\Bigr\}.
\label{eq:m0_p0}
\end{align}

To update \(R=\operatorname{diag}(r_1,\dots,r_p)\), we have
\begin{align}
r_j^{\mathrm{new}}
&=
\frac{1}{\sum_{i=1}^n T_i}
\sum_{i=1}^n\sum_{t=1}^{T_i}
\Bigg[
D_{it,j}^2
-2D_{it,j}\,\bar H_{i,j\cdot}\widehat m_{it}
\nonumber\\
&\hspace{2.2cm}
+
\operatorname{tr}\!\left\{
S_H^{\top} E_j^{\top}\widehat Q_{it}E_j S_H
\bigl(\Omega_{h,i}+\nu_{h,i}\nu_{h,i}^{\top}\bigr)
\right\}
\Bigg],
\label{eq:Rj_update_final}
\end{align}
where \(\bar H_i\) is defined by $\operatorname{vec}(\bar H_i)=S_H\nu_{h,i}$,
and \(E_j\) is the row-selector matrix such that $H_{i,j\cdot}=E_j S_H h_i.$

\section{Additional simulation for partial anchoring under localized random effects}
\label{app:sim_partial_anchor_localized}

To examine the behavior of partial anchoring, we consider an additional simulation setting in which one subject-specific random effect is informed by the entire trajectory, while a second random effect is informed only by an early segment of the trajectory. This setting is designed to favor partial anchored variational inference, since the posterior uncertainty of the localized random effect may remain non-negligible even for long trajectories.

For the model setup, consider the Gaussian MHMM setting with $p=d=1$. To investigate the behavior of partial anchoring, we modify the emission model by
\[
Y_{it}\mid (U_{it}=k,f_i^{(a)},f_i^{(b)})
\sim
N\!\left(
\mu_k + f_i^{(a)} + f_i^{(b)} \mathbf{1}(t\le t_0),\,
\sigma_k^2
\right),
\]
where $f_i^{(a)} \sim N(0,\tau_a^2)$ and $f_i^{(b)} \sim N(0,\tau_b^2)$ are independent subject-specific random effects.
In this model, \(f_i^{(a)}\) is a global random intercept that affects all time points, whereas \(f_i^{(b)}\) is a localized random effect that affects only the first \(t_0\) observations. 
We set \(K=2\), with state-specific variances \(\sigma_1^2=\sigma_2^2=1\) and random-effect variances \(\tau_a^2=1\) and \(\tau_b^2=1.5\). The transition matrix $\Gamma$ and the initial distribution $\pi$ are taken to be the same as those in Simulation Setting~1 in Section \ref{sec:simulation}.

We consider two choices of the state-specific means: a more challenging setting with weaker state separation, \((\mu_1,\mu_2)=(0.5,-0.5)\), and an easier setting with stronger state separation, \((\mu_1,\mu_2)=(1.5,-1.5)\). We combine these with three values of the localization threshold, $t_0 \in \{5,10,15\}.$
For each setting, we generate \(n=40\) subjects with trajectory length \(T=40\), and compare the AVEM algorithm with the partial-anchor method (PAVEM) using \(5\), \(7\), and \(9\) quadrature nodes for the localized random effect \(f_i^{(b)}\). 
Each configuration is repeated over \(100\) Monte Carlo replicates.

\begin{figure}[t]
    \centering
    \includegraphics[width=\textwidth]{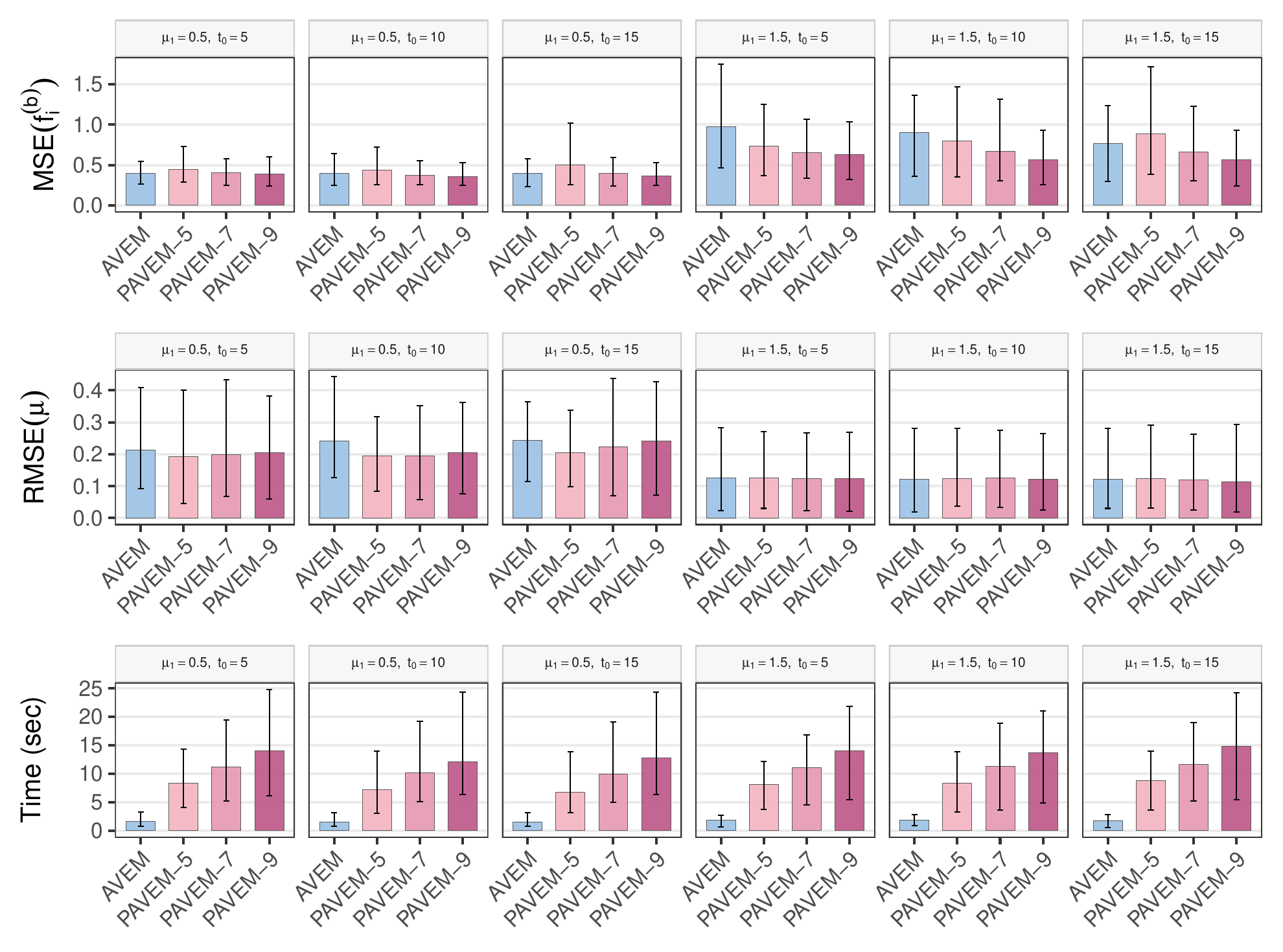}
    \caption{
    Comparison of AVEM and PAVEM in the localized-random-effect setting. The columns correspond to different values of \(\mu_1\) and \(t_0\), and the rows show the estimation error of \(f_i^{(b)}\), the estimation error of \(\mu\), and the total computation time. Error bars represent empirical $90\%$ intervals over simulation replicates
    }
    \label{fig:partial_anchor_sim}
\end{figure}

Figure~\ref{fig:partial_anchor_sim} summarizes the results. The first row reports the averaged MSE for estimating the localized random effect \(f_i^{(b)}\) across 100 replicates, the second row reports the averaged RMSE for estimating the state-specific means \(\mu\), and the third row reports the averaged total computation time. Error bars represent empirical $90\%$ intervals over simulation replicates.

Overall, the results indicate that PAVEM mainly improves estimation of the localized random effect \(f_i^{(b)}\). This improvement is especially clear in the setting with stronger state separation, where the MSE of \(f_i^{(b)}\) decreases substantially as the number of quadrature nodes increases. In the more challenging setting with weaker state separation, the gains for \(f_i^{(b)}\) are more modest but remain visible when the number of nodes is sufficiently large. By contrast, the advantage of PAVEM for the global emission means \(\mu\) is limited and not systematic: in some settings it yields a modest improvement, whereas in many others the methods perform similarly. At the same time, AVEM remains substantially faster in all settings and is broadly competitive for the global model parameters. Therefore, this additional experiment suggests that partial anchoring is most useful when accurate recovery of the localized random effect is of primary interest, while the standard anchored method remains an attractive and computationally efficient default.

\section{Proofs}

\paragraph{Proof of Lemma \ref{lemma1}}
To simplify notation, we suppress the subscript \(i\) on \(f_i\), \(U_i\), \(D_i\), and \(q_i(\cdot) \) throughout the proof.
By direct calculation,
\begin{align*}
&\KL\bigl(q(U,f)\|p(U,f\mid D)\bigr) \\
&=
\int q(U,f)\log\frac{q(U,f)}{p(U,f\mid D)}\,dU\,df \\
&=
\int q(U\mid f)q(f)
\log\frac{q(U\mid f)q(f)}{p(U\mid f,D)\,p(f\mid D)}
\,dU\,df \\
&=
\int q(f)\left(
\int q(U\mid f)\log\frac{q(U\mid f)}{p(U\mid f,D)}\,dU
\right)df
+
\int q(f)\log\frac{q(f)}{p(f \mid D)}
\left(\int q(U\mid f)\,dU\right) df \\
&=
\mathbb{E}_{q(f)}\!\left[
\KL\bigl(q(U\mid f)\|p(U\mid f,D)\bigr)
\right]
+
\KL\bigl(q(f)\|p(f\mid D)\bigr),
\end{align*}
where we used \(\int q(U\mid f)\,dU=1\). Since the second term does not depend on \( U \) and the first term is nonnegative, the minimum is attained when
\[
\KL\bigl(q(U\mid f)\|p(U\mid f,D)\bigr)=0
\quad\text{for } q(f)\text{-a.e. }f,
\]
that is, when \(q(U\mid f)=p(U\mid f,D)\).

\bigskip
\paragraph{Proof of Lemma \ref{lemma2}}
We suppress the subscript \(i\) throughout the proof.
Since \(q(U\mid f)=p(U\mid f_0,D)\), we have
\begin{align*}
\KL\bigl(q(U,f)\|p(U,f\mid D)\bigr)
&=
\int q(f)\,p(U\mid f_0,D)
\log
\frac{p(U\mid f_0,D)\,q(f)}{p(U\mid f,D)\,p(f\mid D)}
\,dU\,df \\
&=
\mathbb E_{q(f)}
\!\left[
\KL\bigl(p(U\mid f_0,D)\|p(U\mid f,D)\bigr)
\right]
+
\KL\bigl(q(f)\|p(f\mid D)\bigr).
\end{align*}
The second term does not depend on \(f_0\), so minimizing over \(f_0\) is equivalent to minimizing
\[
\mathbb E_{q(f)}
\!\left[
\KL\bigl(p(U\mid f_0,D)\|p(U\mid f,D)\bigr)
\right].
\]

\bigskip
\paragraph{Proof of Theorem \ref{thm:mean_anchor_bound}}
Applying Condition~\ref{cond:KL_smooth}, we obtain
\begin{align*}
\mathcal J_i(\bar f_{0i})
&=
\E_{q_i(f_i)}
\!\left[
\KL\!\Bigl(
p(U_i\mid D_i,\bar f_{0i};\theta)
\,\Big\|\,
p(U_i\mid D_i,f_i;\theta)
\Bigr)
\right] \\
&\le
L\, T\,\E_{q_i(f_i)}\|\bar f_{0i}-f_i\|^2 \\
&=
L\, T\,\tr\bigl(\Var_{q_i}(f_i)\bigr).
\end{align*}
On the other hand, since \(f_{0i}^\ast\) minimizes \(\mathcal J_i(f_{0i})\), we have
\[
\mathcal J_i(f_{0i}^\ast)\le \mathcal J_i(\bar f_{0i}).
\]
Therefore,
\[
0
\le
\mathcal J_i(\bar f_{0i})-\mathcal J_i(f_{0i}^\ast)
\le
\mathcal J_i(\bar f_{0i})
\le
L\, T\,\tr\bigl(\Var_{q_i}(f_i)\bigr).
\]
This completes the proof.

\bigskip
\paragraph{Proof of Theorem \ref{thm:approx_monotonicity}}
By Proposition~\ref{prop:fixed_anchor_monotonicity}, for the anchor \(f_0^{(m)}\) fixed at iteration \(m\),
\begin{equation}
\bar{\mathcal L}_A\!\left(f_0^{(m)},q^{(m+1)},\theta^{(m+1)}\right)
\ge
\bar{\mathcal L}_A\!\left(f_0^{(m)},q^{(m)},\theta^{(m)}\right).
\label{eq:fixed_anchor_mono}
\end{equation}
Thus, it remains to control the effect of refreshing the anchor from \(f_0^{(m)}\) to \(f_0^{(m+1)}\).

Using Bayes' rule, the anchored ELBO in \eqref{equ:anchor_elbo} can be rewritten as
\begin{align*}
\mathcal L_A(f_0,q,\theta)
&=
\sum_{i=1}^n
\mathbb E_{q_i(f_i)p(U_i\mid D_i,f_{0i};\theta)}
\Bigl[
\log p(U_i\mid D_i,f_i;\theta)
-
\log p(U_i\mid D_i,f_{0i};\theta)
\Bigr]
\\
&\qquad
+
\sum_{i=1}^n
\mathbb E_{q_i(f_i)}
\Bigl[
\log p(D_i,f_i;\theta)-\log q_i(f_i)
\Bigr] \\
&=
-\sum_{i=1}^n
\mathbb E_{q_i(f_i)}
\left[
\KL\!\Bigl(
p(U_i\mid D_i,f_{0i};\theta)
\,\Big\|\,
p(U_i\mid D_i,f_i;\theta)
\Bigr)
\right]
+
C(q,\theta),
\end{align*}
where
\[
C(q,\theta)
:=
\sum_{i=1}^n
\mathbb E_{q_i(f_i)}
\Bigl[
\log p(D_i,f_i;\theta)-\log q_i(f_i)
\Bigr]
\]
does not depend on \(f_0\). 
Denote
\[
\mathcal J_i^{(m+1)}(f_{0i})
:=
\mathbb E_{q_i^{(m+1)}(f_i)}
\left[
\KL\!\Bigl(
p(U_i\mid D_i,f_{0i};\theta^{(m+1)})
\,\Big\|\,
p(U_i\mid D_i,f_i;\theta^{(m+1)})
\Bigr)
\right].
\]

Hence,
\begin{align}
&\bar{\mathcal L}_A\!\left(f_0^{(m+1)},q^{(m+1)},\theta^{(m+1)}\right)
-
\bar{\mathcal L}_A\!\left(f_0^{(m)},q^{(m+1)},\theta^{(m+1)}\right)
\nonumber\\
&\qquad=
-\frac{1}{nT}\sum_{i=1}^n
\Bigl\{
\mathcal J_i^{(m+1)}\!\left(f_{0i}^{(m+1)}\right)
-
\mathcal J_i^{(m+1)}\!\left(f_{0i}^{(m)}\right)
\Bigr\},
\label{eq:anchor_refresh_identity}
\end{align}

Now \(f_{0i}^{(m+1)}\) is chosen as the mean-anchor under \(q_i^{(m+1)}(f_i)\). By Corollary~\ref{cor:mean_anchor_rate},
\[
\mathcal J_i^{(m+1)}\left(f_{0i}^{(m+1)}\right)-\mathcal J_i^{(m+1)}\left(f_{0i}^{\ast(m+1)}\right)
\le
 L L' T^{1-r},
\]
where \(f_{0i}^{\ast(m+1)}\) denotes the optimal-anchor for subject \(i\) at iteration \(m+1\). Since \(f_{0i}^{\ast(m+1)}\) minimizes \(\mathcal J_i^{(m+1)}\), we also have
\[
\mathcal J_i^{(m+1)}\left(f_{0i}^{\ast(m+1)}\right)
\le
\mathcal J_i^{(m+1)}\left(f_{0i}^{(m)}\right).
\]
Therefore,
\begin{align*}
\mathcal J_i^{(m+1)}\left(f_{0i}^{(m+1)}\right)-\mathcal J_i^{(m+1)}\left(f_{0i}^{(m)}\right)
&=
\Bigl\{
\mathcal J_i^{(m+1)}\left(f_{0i}^{(m+1)}\right)-\mathcal J_i^{(m+1)}\left(f_{0i}^{\ast(m+1)}\right)
\Bigr\} \\
& \qquad +
\Bigl\{
\mathcal J_i^{(m+1)}\left(f_{0i}^{\ast(m+1)}\right)-\mathcal J_i^{(m+1)}\left(f_{0i}^{(m)}\right)
\Bigr\}
\\
&\le
\mathcal J_i^{(m+1)}\left(f_{0i}^{(m+1)}\right)-\mathcal J_i^{(m+1)}\left(f_{0i}^{\ast(m+1)}\right)
\\
&\le
 L L' T^{1-r}.
\end{align*}
Thus
\[
\frac{1}{nT}\sum_{i=1}^n
\Bigl\{
\mathcal J_i^{(m+1)}\left(f_{0i}^{(m+1)}\right)-\mathcal J_i^{(m+1)}\left(f_{0i}^{(m)}\right)
\Bigr\}
\le  L L' T^{-r}.
\]
Substituting this into \eqref{eq:anchor_refresh_identity} yields
\begin{equation}
\bar{\mathcal L}_A\!\left(f_0^{(m+1)},q^{(m+1)},\theta^{(m+1)}\right)
\ge
\bar{\mathcal L}_A\!\left(f_0^{(m)},q^{(m+1)},\theta^{(m+1)}\right)
-
L L' T^{-r}.
\label{eq:anchor_refresh_bound}
\end{equation}
Combining \eqref{eq:fixed_anchor_mono} and \eqref{eq:anchor_refresh_bound}, we conclude that
\[
\bar{\mathcal L}_A\!\left(f_0^{(m+1)},q^{(m+1)},\theta^{(m+1)}\right)
\ge
\bar{\mathcal L}_A\!\left(f_0^{(m)},q^{(m)},\theta^{(m)}\right)
-
L L' T^{-r},
\]
which proves the theorem.

\bigskip
\paragraph{Proof of Theorem \ref{thm:elbo_gap}}
As shown in the proof of Theorem~\ref{thm:approx_monotonicity}, the anchored ELBO can be written as
\[
\mathcal L_A(f_0,q,\theta)
=
-\sum_{i=1}^n \mathcal J_i(f_{0i})
+
C(q,\theta),
\]
where
\[
\mathcal J_i(f_{0i})
:=
\mathbb E_{q_i(f_i)}
\!\left[
\KL\!\Bigl(
p(U_i\mid D_i,f_{0i};\theta)
\,\Big\|\,
p(U_i\mid D_i,f_i;\theta)
\Bigr)
\right],
\]
and
\[
C(q,\theta)
:=
\sum_{i=1}^n
\mathbb E_{q_i(f_i)}
\Bigl[
\log p(D_i,f_i;\theta)-\log q_i(f_i)
\Bigr]
\]
does not depend on \(f_0\).

On the other hand, under the original structured variational family
\[
q_i(U_i,f_i)=p(U_i\mid D_i,f_i;\theta)\,q_i(f_i),
\]
the corresponding ELBO satisfies
\begin{align*}
\mathcal L_S(q,\theta)
&=
\sum_{i=1}^n
\mathbb E_{q_i(f_i)p(U_i\mid D_i,f_i;\theta)}
\Bigl[
\log p(D_i,U_i,f_i;\theta)
-
\log p(U_i\mid D_i,f_i;\theta)
-
\log q_i(f_i)
\Bigr]
\\
&=
\sum_{i=1}^n
\mathbb E_{q_i(f_i)}
\Bigl[
\log p(D_i,f_i;\theta)-\log q_i(f_i)
\Bigr]
\\
&=
C(q,\theta).
\end{align*}
Therefore,
\[
\bar{\mathcal L}_S(q,\theta)-\bar{\mathcal L}_A(f_0,q,\theta)
=
\frac{1}{nT}\sum_{i=1}^n \mathcal J_i(f_{0i}) \ge 0.
\]

Now choose the anchor as the mean-anchor $\bar f_{0i} =\mathbb E_{q_i(f_i)}(f_i)$.
By Condition~\ref{cond:KL_smooth},
\[
\mathcal J_i(\bar f_{0i})
\le
L\, T\, \mathbb E_{q_i(f_i)}\|\bar f_{0i}-f_i\|^2
=
L\, T\, \tr\bigl(\Var_{q_i}(f_i)\bigr).
\]
Condition~\ref{cond:var_concentration} yields
\[
\max_{1\le i\le n}\tr\bigl(\Var_{q_i}(f_i)\bigr) \le L' T^{-r},
\]
and hence
\[
0 \le \bar{\mathcal L}_S(q,\theta)-\bar{\mathcal L}_A(f_0,q,\theta) = \frac{1}{nT}\sum_{i=1}^n \mathcal J_i(f_{0i}) \le L L' T^{-r},
\]
which proves the theorem.

\vskip 0.2in
\bibliography{references}

@article{baum1970maximization,
  title={A maximization technique occurring in the statistical analysis of probabilistic functions of {M}arkov chains},
  author={Baum, Leonard E and Petrie, Ted and Soules, George and Weiss, Norman},
  journal={The Annals of Mathematical Statistics},
  volume={41},
  number={1},
  pages={164--171},
  year={1970},
  publisher={JSTOR}
}

@article{dempster1977maximum,
  title={Maximum likelihood from incomplete data via the {EM} algorithm},
  author={Dempster, Arthur P and Laird, Nan M and Rubin, Donald B},
  journal={Journal of the Royal Statistical Society: Series B (Methodological)},
  volume={39},
  number={1},
  pages={1--22},
  year={1977},
  publisher={Wiley Online Library}
}

@incollection{neal1998view,
  title={A view of the EM algorithm that justifies incremental, sparse, and other variants},
  author={Neal, Radford M and Hinton, Geoffrey E},
  booktitle={Learning in graphical models},
  pages={355--368},
  year={1998},
  publisher={Springer}
}

@article{jordan1999introduction,
  title={An introduction to variational methods for graphical models},
  author={Jordan, Michael I and Ghahramani, Zoubin and Jaakkola, Tommi S and Saul, Lawrence K},
  journal={Machine learning},
  volume={37},
  number={2},
  pages={183--233},
  year={1999},
  publisher={Springer}
}

@article{altman2007mixed,
  title={Mixed hidden Markov models: An extension of the hidden Markov model to the longitudinal data setting},
  author={Altman, Rachel MacKay},
  journal={Journal of the American Statistical Association},
  volume={102},
  number={477},
  pages={201--210},
  year={2007},
  publisher={Taylor \& Francis}
}

@inproceedings{ranganath2016hierarchical,
  title={Hierarchical variational models},
  author={Ranganath, Rajesh and Tran, Dustin and Blei, David},
  booktitle={The 33rd International Conference on Machine Learning},
  year={2016}
}

@article{blei2017variational,
  title={Variational inference: A review for statisticians},
  author={Blei, David M and Kucukelbir, Alp and McAuliffe, Jon D},
  journal={Journal of the American statistical Association},
  volume={112},
  number={518},
  pages={859--877},
  year={2017},
  publisher={Taylor \& Francis}
}

@article{ashwood2022mice,
  title={Mice alternate between discrete strategies during perceptual decision-making},
  author={Ashwood, Zoe C and Roy, Nicholas A and Stone, Iris R and International Brain Laboratory and Urai, Anne E and Churchland, Anne K and Pouget, Alexandre and Pillow, Jonathan W},
  journal={Nature Neuroscience},
  volume={25},
  number={2},
  pages={201--212},
  year={2022},
  publisher={Nature Publishing Group US New York}
}

@article{bian2025ddm,
  title={Joint Modeling for Learning Decision-Making Dynamics in Behavioral Experiments},
  author={Bian, Yuan and Guo, Xingche and Wang, Yuanjia},
  journal={The Annals of Applied Statisitcs},
  volume={19},
  number={4},
  year={2025},
  pages={3372--3393},
}

@article{guo2025hmm,
  title={{HMM} for discovering decision-making dynamics using reinforcement learning experiments},
  author={Guo, Xingche and Zeng, Donglin and Wang, Yuanjia},
  journal={Biostatistics},
  volume={26},
  number={1},
  year={2025},
  pages={Article kxae033}
}

@article{rabiner1989tutorial,
  title={A tutorial on hidden Markov models and selected applications in speech recognition},
  author={Rabiner, Lawrence R.},
  journal={Proceedings of the IEEE},
  volume={77},
  number={2},
  pages={257--286},
  year={1989}
}

@book{cappe2005inference,
  title={Inference in hidden Markov models},
  author={Capp{\'e}, Olivier and Moulines, Eric and Ryd{\'e}n, Tobias},
  year={2005},
  publisher={Springer}
}

@article{shumway1982approach,
  title={An approach to time series smoothing and forecasting using the EM algorithm},
  author={Shumway, Robert H and Stoffer, David S},
  journal={Journal of time series analysis},
  volume={3},
  number={4},
  pages={253--264},
  year={1982},
  publisher={Wiley Online Library}
}

@article{maruotti2011mixed,
  title={Mixed hidden Markov models for longitudinal data: An overview},
  author={Maruotti, Antonello},
  journal={International Statistical Review},
  volume={79},
  number={3},
  pages={427--454},
  year={2011},
  publisher={Wiley Online Library}
}

@article{ormerod2010explaining,
  title={Explaining variational approximations},
  author={Ormerod, John T and Wand, Matt P},
  journal={The American Statistician},
  volume={64},
  number={2},
  pages={140--153},
  year={2010},
  publisher={Taylor \& Francis}
}

@article{saul1995exploiting,
  title={Exploiting tractable substructures in intractable networks},
  author={Saul, Lawrence and Jordan, Michael},
  journal={Advances in neural information processing systems},
  volume={8},
  year={1995}
}

@article{ghahramani2000variational,
  title={Variational learning for switching state-space models},
  author={Ghahramani, Zoubin and Hinton, Geoffrey E.},
  journal={Neural Computation},
  volume={12},
  number={4},
  pages={831--864},
  year={2000}
}

@article{mcclintock2021worth,
  title={Worth the effort? A practical examination of random effects in hidden Markov models for animal telemetry data},
  author={McClintock, Brett T},
  journal={Methods in Ecology and Evolution},
  volume={12},
  number={8},
  pages={1475--1497},
  year={2021},
  publisher={Wiley Online Library}
}

@article{guo2024hierarchical,
  title={A hierarchical random effects state-space model for modeling brain activities from electroencephalogram data},
  author={Guo, Xingche and Yang, Bin and Loh, Ji Meng and Wang, Qinxia and Wang, Yuanjia},
  journal={Biometrics},
  volume={80},
  number={4},
  pages={ujae130},
  year={2024},
  publisher={Oxford University Press}
}

@article{kappe2018random,
  title={A random coefficients mixture hidden Markov model for marketing research},
  author={Kappe, Eelco and Blank, Ashley Stadler and DeSarbo, Wayne S},
  journal={International Journal of Research in Marketing},
  volume={35},
  number={3},
  pages={415--431},
  year={2018},
  publisher={Elsevier}
}

@article{welch1995introduction,
  title={An introduction to the Kalman filter},
  author={Welch, Greg and Bishop, Gary and others},
  year={1995},
  publisher={Chapel Hill, NC, USA}
}

@article{zhang2018advances,
  title={Advances in variational inference},
  author={Zhang, Cheng and B{\"u}tepage, Judith and Kjellstr{\"o}m, Hedvig and Mandt, Stephan},
  journal={IEEE transactions on pattern analysis and machine intelligence},
  volume={41},
  number={8},
  pages={2008--2026},
  year={2018},
  publisher={IEEE}
}

@article{coviello2014clustering,
  title={Clustering hidden Markov models with variational HEM},
  author={Coviello, Emanuele and Chan, Antoni B and Lanckriet, Gert RG},
  journal={The Journal of Machine Learning Research},
  volume={15},
  number={1},
  pages={697--747},
  year={2014},
  publisher={JMLR. org}
}

@article{wang2004lack,
  title={Lack of consistency of mean field and variational Bayes approximations for state space models},
  author={Wang, Bo and Titterington, DM},
  journal={Neural Processing Letters},
  volume={20},
  number={3},
  pages={151--170},
  year={2004},
  publisher={Springer}
}

@article{liu2011mixed,
  title={Mixed-effects state-space models for analysis of longitudinal dynamic systems},
  author={Liu, Dacheng and Lu, Tao and Niu, Xu-Feng and Wu, Hulin},
  journal={Biometrics},
  volume={67},
  number={2},
  pages={476--485},
  year={2011},
  publisher={Oxford University Press}
}

@article{wang2023latent,
  title={A latent state space model for estimating brain dynamics from electroencephalogram (EEG) data},
  author={Wang, Qinxia and Loh, Ji Meng and He, Xiaofu and Wang, Yuanjia},
  journal={Biometrics},
  volume={79},
  number={3},
  pages={2444--2457},
  year={2023},
  publisher={Wiley Online Library}
}

@article{foti2014stochastic,
  title={Stochastic variational inference for hidden Markov models},
  author={Foti, Nicholas J and Xu, Jason and Laird, Dillon and Fox, Emily B},
  journal={Advances in neural information processing systems},
  volume={27},
  year={2014}
}

@inproceedings{xing2003generalized,
  title={A generalized mean field algorithm for variational inference in exponential families},
  author={Xing, Eric P. and Jordan, Michael I. and Russell, Stuart},
  booktitle={Proceedings of the Nineteenth Conference on Uncertainty in Artificial Intelligence},
  pages={583--591},
  year={2003}
}

@article{wu2026extending,
  title={Extending mean-field variational inference via entropic regularization: theory and computation},
  author={Wu, Bohan and Blei, David M},
  journal={Journal of Machine Learning Research},
  volume={27},
  number={7},
  pages={1--68},
  year={2026}
}

@article{yu2010hidden,
  title={Hidden semi-Markov models},
  author={Yu, Shun-Zheng},
  journal={Artificial intelligence},
  volume={174},
  number={2},
  pages={215--243},
  year={2010},
  publisher={Elsevier}
}

@article{hamilton1989new,
  title={A new approach to the economic analysis of nonstationary time series and the business cycle},
  author={Hamilton, James D},
  journal={Econometrica: Journal of the econometric society},
  pages={357--384},
  year={1989},
  publisher={JSTOR}
}

@article{bertsekas1997nonlinear,
  title={Nonlinear programming},
  author={Bertsekas, Dimitri P},
  journal={Journal of the Operational Research Society},
  volume={48},
  number={3},
  pages={334--334},
  year={1997},
  publisher={Taylor \& Francis}
}

@inproceedings{bottou2010large,
  title={Large-scale machine learning with stochastic gradient descent},
  author={Bottou, L{\'e}on},
  booktitle={Proceedings of COMPSTAT'2010: 19th International Conference on Computational StatisticsParis France, August 22-27, 2010 Keynote, Invited and Contributed Papers},
  pages={177--186},
  year={2010},
  organization={Springer}
}

@inproceedings{ranganath2014black,
  title={Black box variational inference},
  author={Ranganath, Rajesh and Gerrish, Sean and Blei, David},
  booktitle={Artificial intelligence and statistics},
  pages={814--822},
  year={2014},
  organization={PMLR}
}

\end{document}